\documentclass[11pt, letter]{amsart}
\usepackage{amsthm,amsmath,amsxtra,amscd,amssymb,xypic,color,xspace,esint}
\usepackage{latexsym,amsmath,amssymb,mathrsfs}
\usepackage{amsfonts,eucal,amsthm,graphicx,color}

\usepackage{graphicx} 
\usepackage{hyperref}

\hypersetup{
    colorlinks = true,
   linktocpage = true,
    citecolor = {blue}
}

\usepackage{algpseudocode}
\usepackage{subcaption}
\usepackage{svg}

\usepackage{latexsym}
\usepackage{amsmath}
\usepackage{amssymb}
\usepackage{mathrsfs}
\usepackage{amsfonts}
\usepackage{eucal}
\usepackage{amsthm}
\usepackage{graphicx}
\usepackage{color}
\usepackage{youngtab}
\usepackage{slashed}
\usepackage{tikz-cd}

\newcommand {\qe}{\mathfrak{q}}
\newcommand {\pe}{\mathfrak{p}}

\newcommand {\CalC} {\mathcal C}

\newcommand {\CalD} {\mathcal D}
\newcommand {\CalE} {\mathcal E}

\newcommand {\CalL} {\mathcal L}
\newcommand {\CalM} {\mathcal M}
\newcommand {\CalN} {\mathcal N}
\newcommand {\CalO} {\mathcal O}

\newcommand {\CalQ} {\mathcal Q}
\newcommand {\CalR} {\mathcal R}

\newcommand {\CalT} {\mathcal T}

\newcommand {\CalX} {\mathcal X}

\newcommand {\CalZ} {\mathcal Z}

\newcommand {\eig} {\mathscr Z}

\newcommand {\BC}   {\mathbb C}

\newcommand {\BR}   {\mathbb R}
\newcommand {\BP}   {\mathbb P}

\newcommand {\ba}  {\mathbf{a}}

\newcommand {\bA}{ \mathbf{A}}
\newcommand {\bZ}{ \mathbf{Z}}

\newcommand {\bt}   {\mathbf t}

\newcommand {\BZ}   {\mathbb Z}

\newcommand {\sa}{\sf a}
\newcommand {\si}{\sf i}
\newcommand {\sj}{\sf j}

\newcommand {\ve}{\varepsilon}

\newcommand {\vt}{\vartheta}

\newcommand{\bla}{\boldsymbol{\lambda}}
\newcommand{\bve}{\boldsymbol{\ve}}
\newcommand{\bmu}{\boldsymbol{\mu}}

\newcommand{\bq}{\boldsymbol{\qe}}

\newcommand{\ii}{\mathrm{i}}

\newcommand{\beq}{\begin{equation}}
\newcommand{\eeq}{\end{equation}}

\begin{document}

\title[Vershik-Kerov Higher times]{Vershik-Kerov\\ in higher times}
\author{Andrei Grekov, Nikita Nekrasov}
\address{Simons Center for Geometry and Physics$^{n}$\\
Yang Institute for Theoretical Physics$^{g,n}$\\ 
Stony Brook University, Stony Brook NY 11794-3636, USA}

\begin{abstract}
Several generalizations of Vershik-Kerov limit shape problem are motivated by topological string theory and supersymmetric gauge theory instanton count. In this paper specifically we study the ${\hat A}_r$- and $A_r$-quiver type models (circular and linear quiver theories). We also briefly discuss
the double-elliptic generalization of the Vershik-Kerov problem, related to six dimensional 
${\CalN}=2$ gauge theory compactified on a torus, and to elliptic cohomology of the Hilbert scheme of points on ${\BC}^{2}$. We prove that the limit shape in that setting is governed by a genus two algebraic curve, suggesting unexpected dualities between the enumerative and equivariant parameters. \\

In loving memory of Anatoly Moiseevich Vershik (1933-2024), with gratitude and admiration. Great mathematician and brave man, who lived through all kinds of historical periods, may he live in higher space and times. 
\end{abstract}

\maketitle
\section{Introduction}

$\bullet$ \emph{Partitions, arms, legs, hooks, contents and limit shapes.}

{}
Recall that a partition ${\lambda}$ of a non-negative integer $N = |{\lambda}|$ is a non-increasing sequence 
\beq
\begin{aligned}
& {\lambda} = \left({\lambda}_{1} , {\lambda}_{2} \, , \,  \ldots \, , \, {\lambda}_{{\ell}({\lambda})}\right)  \, , \\
&  {\lambda}_{i} \in {\BZ}\, , \ i = 1 , \ldots, {\ell}({\lambda})  
 \\
& {\lambda}_{1} \geq {\lambda}_{2} \geq  \ldots \geq {\lambda}_{{\ell}({\lambda})} > 0 \\
\end{aligned}
\eeq
such that
\beq
| {\lambda} | = N  = {\lambda}_{1} + {\lambda}_{2} + \ldots + {\lambda}_{{\ell}({\lambda})} \ . 
\eeq
A pair $(i,j)$ of integers, such that $1 \leq j \leq {\lambda}_{i}$, is called
a box $\square = (i,j) \in \lambda$.  Equivalently, $(i,j) \in \lambda$ iff
$1 \leq i \leq {\lambda}_{j}^{t}$, where $\lambda^t$ is called the \emph{dual or transposed} partition. Obviously $|{\lambda}^t| = |{\lambda}|$, ${\lambda}^{t}_{1} = {\ell}({\lambda})$, ${\lambda}_{1} = {\ell} ({\lambda}^{t})$. 
Given a box $\square$ one defines its \emph{arm-length} $a_{\square}$, \emph{leg-length} $l_{\square}$ and \emph{hook-length} $h_{\square}$ via:
\beq
{\sf a}_{\square} = {\lambda}_{i} - j \, , \ {\sf l}_{\square} = {\lambda}_{j}^{t} - i \, , \
{\sf h}_{\square} = {\sf a}_{\square} + {\sf l}_{\square} + 1 
\label{eq:calh}
\eeq
Given two complex numbers ${\ve}_{1}, {\ve}_{2} \in {\BC}$, the \emph{refined content} of a box ${\square}= (i,j) \in {\lambda}$ is defined
as
\beq
{\sf c}_{\square}({\ve}_{1}, {\ve}_{2}) = {\ve}_{1}(i-1) + {\ve}_{2}(j-1) 
\label{eq:refcont}
\eeq
Given $q_{1}, q_{2} \in {\BC}^{\times}$, the \emph{refined $q$-content} of ${\square} = (i,j)$
is defined as
\beq
{\sf Q}_{\square}(q_{1}, q_{2}) = q_{1}^{i-1} q_{2}^{j-1} 
\label{eq:refqcont}
\eeq
Finally, given an elliptic curve ${\CalE}_{\pe} = {\BC}^{\times}/{\pe}^{\BZ}$, and $q_{1}, q_{2} \in {\CalE}_{\pe}$, the \emph{refined elliptic content} of ${\square} = (i,j)$ is given by the same
formula, but the right hand side of \eqref{eq:refqcont} is now viewed as an element of ${\CalE}_{\pe}$. 

{} An empty partition, the only partition of $N =0$, is denoted by $\emptyset$. 
The only partition ${\lambda} = (1)$ of $N =1$ is sometimes denoted by $\square$. The set of all partitions of $N$ is denoted by $\Lambda_{N}$, $p(N) : = | {\Lambda}_{N} |$. 
Recall Euler's formula
\beq
\frac{1}{{\phi}({\qe})} = \sum_{N=0}^{\infty} p(N) {\qe}^{N}\, , 
\eeq
where
\beq
{\phi}({\qe}) = \prod_{n=1}^{\infty} ( 1- {\qe}^{n} ) 
\label{eq:eulerq}
\eeq

$\bullet$
Fifty years ago, two groups of mathematicians: A.~Vershik (1933-2024) and his student S.~Kerov (1946-2000) \cite{VK} in USSR and B.~Logan and L.~Schepp \cite{LS} in USA, studied the large $N$ asymptotics of the Plancherel measure on the set of irreducible representations $R_{\lambda}$ of symmetric group $S(N)$, equivalently, on  $\Lambda_N$:
\beq
    {\mu}_{N} [\lambda] = \frac{(\text{dim} R_\lambda)^2}{N!} = N! \prod_{\square \in \lambda} \frac{1}{{\sf h}_{\square}^{2}}
    \label{eq:Plancherel}
\eeq
so that
\beq
1 = \sum_{\lambda \in {\Lambda}_{N}} {\mu}_{N} [{\lambda}] 
\eeq
They found that upon rescaling the linear size of $\lambda$ by $\propto \sqrt{N}$, in the $N \to \infty$ limit, the boundary 
 
{}
\centerline{\includegraphics[width=10cm]{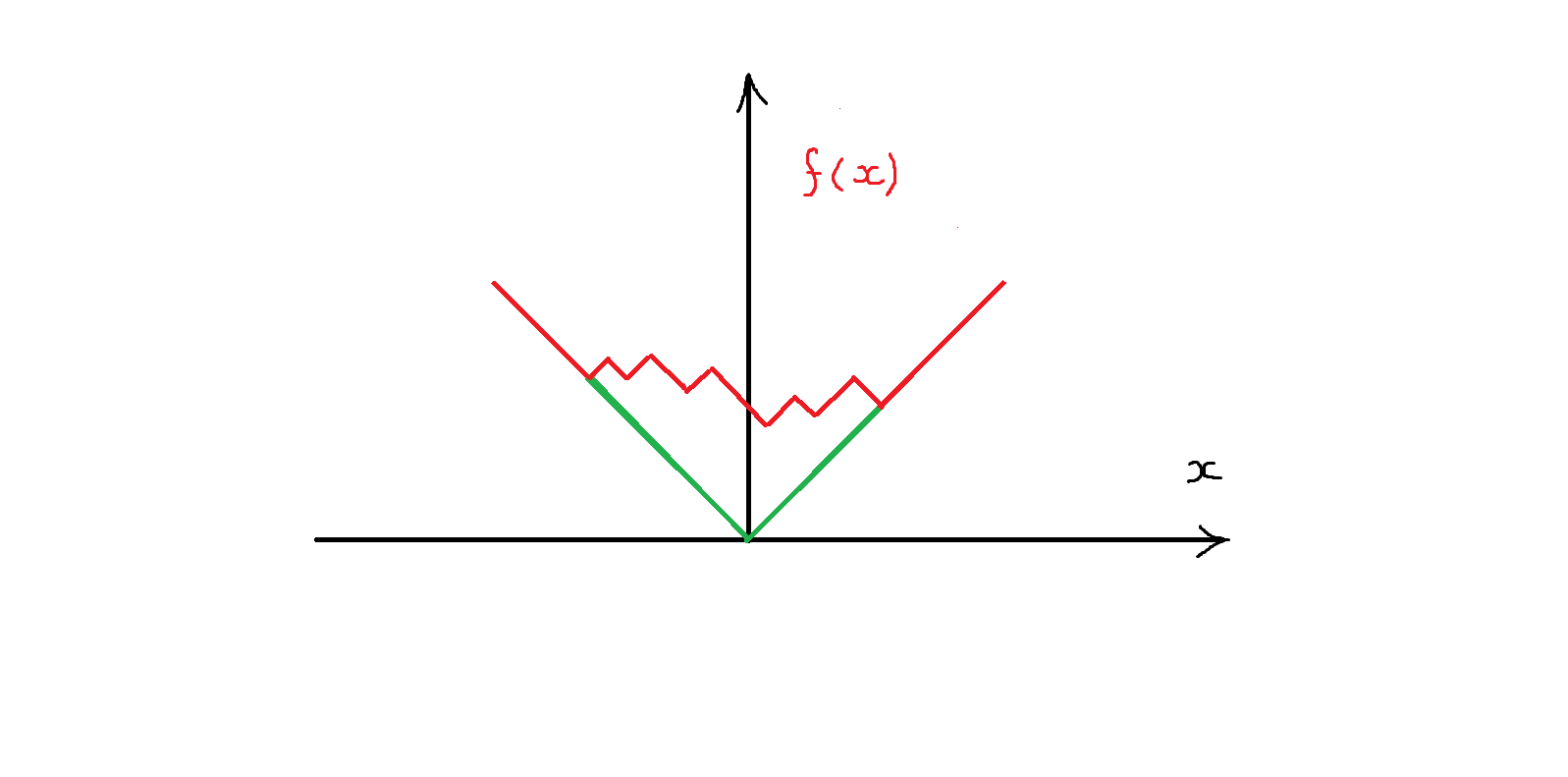}}

{}
\noindent of the set of boxes of $\lambda$ rotated by $135$ degrees approaches
a piecewise smooth curve $f(x)$, called the \emph{the arcsin law}, since the slope $f^{'}(x)$, being equal to ${\rm sign}\left( \frac{x}{2\sqrt{N}} \right)$ for $|x| \geq 2\sqrt{N}$, interpolates between $-1$ and $+1$ as $\frac{2}{\pi} {\rm arcsin}\left( \frac{x}{2\sqrt{N}} \right)$ for $|x| < 2\sqrt{N}$. In a way, large random Young diagrams behave as random $N \times N$ matrices at large $N$. The limit shape problem \cite{VK,LS} (later improved by S.~Kerov into a central limit theorem \cite{K2, K3}) admits a reformulation as a thermodynamic limit of a grand canonical ensemble, defined on the set $\Lambda$ of all partitions $\lambda$:
\beq
{\bmu}_{\Lambda; \hbar} [{\lambda}] = e^{-\frac{{\Lambda}^2}{{\hbar}^2}} \frac{{\Lambda}^{2|{\lambda}|}}{|{\lambda}|!\, {\hbar}^{2|{\lambda}|}}\, {\mu}_{|{\lambda}|} [{\lambda}]
\label{eq:macroplanch}
\eeq
In the thermodynamic limit $\hbar \to 0$ the fugacity is sent to infinity. It follows from Stirling formula that the typical size $|{\lambda}|$ of partitions scales as $\sim 
({\Lambda}/{\hbar})^{2}$. It follows that if the linear size of $\lambda$ is rescaled by $\hbar$, while keeping $\Lambda$ finite, the range of the slope change for the limit shape (support of the second derivative $f^{(2)}$) becomes the interval $(-2{\Lambda}, 2{\Lambda})$. 

$\bullet$ \emph{Instanton measure} The $1$-parametric (the parameter being $({\Lambda}:{\hbar})$)  family of measures \eqref{eq:macroplanch} is at the intersection of two series of generalizations: a ${\hat A}_{r}$ series, with $r \geq 0$,  and a $A_{r}$ series, with $r \geq 1$. 

The $A_{1}$-model is a $4$-parametric family of 
measures (called the \emph{$z$-measures} in \cite{Borodin}), cf. \eqref{eq:refcont}
\begin{multline}
{\bmu}_{\qe; {\ve}_{1}:{\ve}_{2}: m_{1}: m_{2}}^{A_{1}} [{\lambda}] = (1-{\qe})^{-\frac{m_{+}m_{-}}{{\ve}_{1}{\ve}_{2}}}\, {\qe}^{|{\lambda}|} \times \\
 \, \prod_{{\square}\in {\lambda}} \frac{\left(-m_{+}+{\sf c}_{\square}({\ve}_{1}, {\ve}_{2}) \right)\left(m_{-}-{\sf c}_{\square}({\ve}_{1}, {\ve}_{2}) \right)}{\left( -{\ve}_{1}{\sf l}_{\square} + {\ve}_{2}({\sf a}_{\square}+1) \right) \left( {\ve}_{1}({\sf l}_{\square}+1) - {\ve}_{2}{\sf a}_{\square} \right)}
\label{eq:macroa1}
\end{multline}

It is easy to see, that
\beq
{\bmu}_{\qe; {\ve}_{1}:{\ve}_{2}: m_{+}: m_{-}}^{{A}_{1}}[{\lambda}] = {\bmu}_{\qe; {\ve}_{1}:{\ve}_{2}: m_{-}: m_{+}}^{{A}_{1}}[{\lambda}] = 
{\bmu}_{\qe; {\ve}_{2}:{\ve}_{1}: m_{+}: m_{-}}^{{A}_{1}}[{\lambda}^{t}]
\eeq
The normalization
\beq
\sum_{\lambda \in {\Lambda}} {\bmu}_{\qe; {\ve}_{1}:{\ve}_{2}: m_{+}: m_{-}}^{{A}_{1}}[{\lambda}] = 1
\eeq
can be proven either using the Cauchy identity \cite{McD}, or the techniques reviewed below. 

{}
The limit to \eqref{eq:macroplanch} is to set ${\ve}_{1} = - {\ve}_{2} = {\hbar}$, and to send $m_{+}, m_{-} \to \infty$, ${\qe} \to 0$, while keeping
\beq
{\Lambda}^{2}\, = \, {\qe}\, m_{+} m_{-}
\eeq
finite. 

{}
The ${\hat A}_{0}$-model is a $3$-parametric family of measures
\begin{multline}
{\bmu}_{\qe; {\ve}_{1}:{\ve}_{2}: {\ve}_{3}: {\ve}_{4}}^{{\hat A}_{0}} [{\lambda}] = {\phi}({\qe})^{\frac{({\ve}_{1}+{\ve}_{3})({\ve}_{2}+{\ve}_{3})}{{\ve}_{1}{\ve}_{2}}} \, {\qe}^{|{\lambda}|} \, \times \\
\prod_{{\square}\in {\lambda}} \frac{\left( -{\ve}_{1}{\sf l}_{\square} + {\ve}_{2}({\sf a}_{\square}+1)  + {\ve}_{3} \right) \left( {\ve}_{1}({\sf l}_{\square}+1) - {\ve}_{2}{\sf a}_{\square} + {\ve}_{3} \right)}{\left( -{\ve}_{1}{\sf l}_{\square} + {\ve}_{2}({\sf a}_{\square}+1)  \right) \left( {\ve}_{1}({\sf l}_{\square}+1) - {\ve}_{2}{\sf a}_{\square} \right)}
\label{eq:macroah0}
\end{multline}
where 
\beq
{\ve}_{1}+{\ve}_{2}+{\ve}_{3}+{\ve}_{4} = 0
\label{eq:sl4}
\eeq
implying 
\beq
{\bmu}_{\qe; {\ve}_{1}:{\ve}_{2}: {\ve}_{3}: {\ve}_{4}}^{{\hat A}_{0}}[{\lambda}] = {\bmu}_{\qe; {\ve}_{1}:{\ve}_{2}: {\ve}_{4}: {\ve}_{3}}^{{\hat A}_{0}}[{\lambda}] = 
{\bmu}_{\qe; {\ve}_{2}:{\ve}_{1}: {\ve}_{3}: {\ve}_{4}}^{{\hat A}_{0}}[{\lambda}^{t}]
\eeq
The normalization 
\beq
\sum_{\lambda \in \Lambda} {\bmu}_{\qe; {\ve}_{1}:{\ve}_{2}: {\ve}_{3}: {\ve}_{4}}^{{\hat A}_{0}}[{\lambda}]=1
\eeq
is proven in \cite{BPSCFT1} (it would be interesting to derive it from Cauchy identity as well \cite{McD}).

{}
The limit to \eqref{eq:macroplanch} is to set ${\ve}_{1} = - {\ve}_{2} = {\hbar}$, and to send ${\ve}_{3} \to \infty$, ${\qe} \to 0$, while keeping
\beq
{\Lambda}^{2} \, = \, - {\qe}\, {\ve}_{3}^{2}
\eeq
finite. 

$\bullet$ \emph{Expectation values}

An \emph{observable} is a function ${\CalO}$ on 
\beq
\Lambda = \amalg_{N\geq 0} \ {\Lambda}_{N}\, , \ {\CalO}: {\lambda} \mapsto {\CalO}[{\lambda}]
\eeq
For observable $\CalO$ we
define its \emph{expectation value} as
\beq
\langle {\CalO} \rangle_{\qe; {\ve}_{1}:{\ve}_{2}: m_{+}: m_{-}}^{A_{1}} = \sum_{\lambda \in {\Lambda}}\, {\bmu}_{\qe; {\ve}_{1}:{\ve}_{2}: m_{+}: m_{-}}^{A_{1}} \,  {\CalO}[{\lambda}]
\label{eq:veva1}
\eeq
and
\beq
\langle {\CalO} \rangle_{\qe; {\ve}_{1}:{\ve}_{2}: {\ve}_{3}: {\ve}_{4}}^{{\hat A}_{0}} = \sum_{\lambda \in {\Lambda}}\, {\bmu}_{\qe; {\ve}_{1}:{\ve}_{2}: {\ve}_{3}: {\ve}_{4}}^{{\hat A}_{0}} \,  {\CalO}[{\lambda}]
\label{eq:veva2}
\eeq

$\bullet$ \emph{Limit shapes}
The limit shape problem for these generalizations is to study the asymptotics ${\bmu}_{\ldots}^{\ldots} [{\lambda}]$ when ${\ve}_{1}, {\ve}_{2} \to 0$ while keeping the parameters ${\qe}, m_{+}, m_{-}, {\ve}_{3}$ finite. 

$\bullet$ \emph{$Y$-observables} A useful way to keep track of the shape of $\lambda$ is by studying the expectation values of observables built out of the $Y$-observable, defined by
\beq
Y(x) [ {\lambda} ] \, = \, x \, \prod_{{\square} \in {\lambda}} \frac{(x-c_{\square}({\ve}_{1}, {\ve}_{2})-{\ve}_{1}) (x-c_{\square}({\ve}_{1}, {\ve}_{2})-{\ve}_{2})}{(x-c_{\square}({\ve}_{1}, {\ve}_{2})) (x-c_{\square}({\ve}_{1}, {\ve}_{2})-{\ve})}
\label{eq:yobs}\eeq
where
\beq
{\ve} = {\ve}_{1} + {\ve}_{2}
\eeq
For example 
\beq
Y(x) [{\emptyset}] = x\, , \ Y(x) [{\square}] = \frac{(x-{\ve}_{1})(x-{\ve}_{2})}{x-{\ve}}\, , \ldots
\label{eq:yobs1}\eeq
An equivalent representation for the $Y$-observable is in terms of the addable and removable boxes \cite{BPSCFT1}, 

 \centerline{\includegraphics[width=3cm]{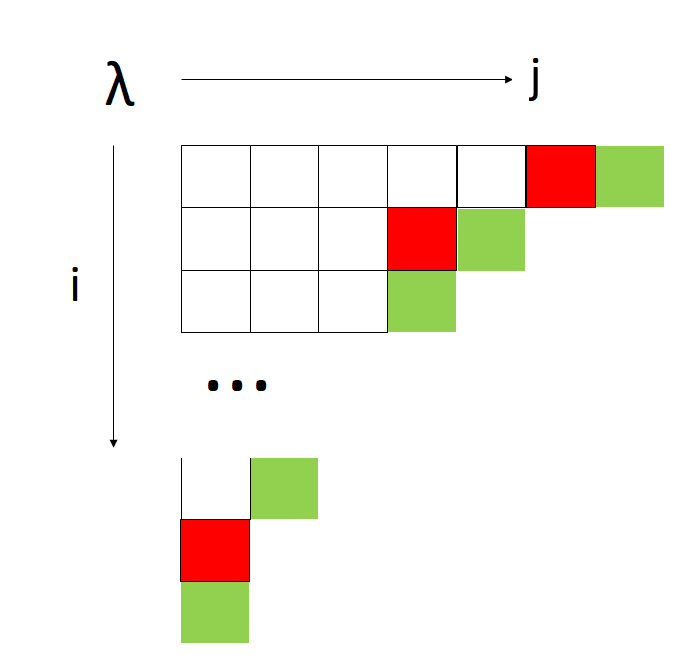}}

 \beq
Y(x) [ {\lambda}] = \frac{\prod_{{\color{green}{\blacksquare}} \in {\partial}_{+}{\lambda}}  \left( x - c_{{\color{green}{\blacksquare}}}({\ve}_{1}, {\ve}_{2}) \right)}{\prod_{{\color{red}{\blacksquare}} \in {\partial}_{+}{\lambda}}  \left( x - {\ve}-c_{{\color{red}{\blacksquare}}}({\ve}_{1}, {\ve}_{2}) \right)}
\label{eq:yobs2}
 \eeq
 
$\bullet$ \emph{$qq$-characters} We see from \eqref{eq:yobs}, \eqref{eq:yobs1}, \eqref{eq:yobs2} that the expectation values of $Y(x)$ or $Y(x')^{-1}$ have poles in $x$ or $x'$. The remarkable fact \cite{BPSCFT1} about the measures \eqref{eq:macroa1},\eqref{eq:macroah0} is that there is a combination of $Y$'s and $Y^{-1}$'s whose expectation value has no poles in $x$ whatsoever:
the simplest such combination, called a \emph{fundamental $qq$-character} is, for $A_1$-model (note a somewhat different normalization compared to \cite{BPSCFT1}):
\beq
{\CalX}^{A_{1}}(x)= Y(x+{\ve}) + {\qe} \frac{(x-m_{+})(x-m_{-})}{Y(x)}
\label{eq:xa1}
\eeq
and
\begin{multline}
{\CalX}^{{\hat A}_{0}}(x) \, = \, Y(x+{\ve}) 
 \times \\
\sum_{\nu \in {\Lambda}} {\bmu}_{\qe ; {\ve}_{3}:{\ve}_{4}:{\ve}_{1}:{\ve}_{2}} [ {\nu}] \prod_{{\square} \in {\nu}} \frac{Y(x+{\sf c}_{\square}({\ve}_{3}, {\ve}_{4}) - {\ve}_{3}) Y(x+{\sf c}_{\square}({\ve}_{3}, {\ve}_{4}) - {\ve}_{4})}{Y(x+{\sf c}_{\square}({\ve}_{3}, {\ve}_{4}) + {\ve}) Y(x+{\sf c}_{\square}({\ve}_{3}, {\ve}_{4}))}
\label{eq:xa0}
\end{multline}
Note the change of order of the $\ve$-parameters in the ${\bmu}_{\qe ; {\ve}_{3}:{\ve}_{4}:{\ve}_{1}:{\ve}_{2}} [ {\nu}]$ measure in the sum over the \emph{auxiliary}
partition $\nu$ in \eqref{eq:xa0}. 

$\bullet$ \emph{Non-perturbative Dyson-Schwinger equations} 
Using the large $x$ asymptotics 
\beq
Y(x)[{\lambda}] \to x + \frac{{\ve}_{1}{\ve}_{2} |{\lambda}|}{x} + o(x^{-1}) \, , \ x \to \infty
\label{eq:asympty}
\eeq
and the absence of poles theorem \cite{BPSCFT1, BPSCFT2}, one concludes:
\beq
\langle {\CalX}^{A_{1}}(x) \rangle_{\qe; {\ve}_{1}:{\ve}_{2}: m_{+}: m_{-}}^{A_{1}}  = x + u_{1}
\eeq
\beq
\langle {\CalX}^{{\hat A}_{0}}(x) \rangle_{\qe; {\ve}_{1}:{\ve}_{2}: {\ve}_{3}: {\ve}_{4}}^{{\hat A}_{0}} = x + {\hat u}_{0}
\label{eq:veva1}
\eeq
with $u_{1}, {\hat u}_{0}$ some constants in $x$, computable with the help of \eqref{eq:asympty} and other tricks \cite{BPSCFT1}. The DS equations are obtained by expanding ${\CalX}(x)$ in $x$ at infinity and setting to zero the expectation values of the observables, defined as the coefficients of the above expansion at the negative powers of $x$. 

$\bullet$ \emph{Seiberg-Witten curves} The limit shape problem is simplified enormously by the following statement (standard, but not proven in this paper): expectation values of $Y$-observables factorize in the ${\ve}_{1}, {\ve}_{2} \to 0$ limit. Thus, in this limit, the expectation values
\beq
y(x) = \langle Y(x) \rangle
\label{eq:yvev}
\eeq
obey, in the $A_1$ case:
\beq
y + {\qe} \frac{(x-m_{+})(x-m_{-})}{y} = (1+{\qe})(x + u_{1}) \, , \label{eq:swa1}
\eeq
giving a simple rational curve relating $y(x)$ to $x$. The value of $u_{1}$, 
\beq
u_{1} = -\frac{\qe}{1+{\qe}} (m_{+} +m_{-})
\eeq
can be computed either from the large $x$ expansion \eqref{eq:asympty} in the limit
${\ve}_{1}, {\ve}_{2} \to 0$, or, more
systematically, by computing the period
of the \emph{Seiberg-Witten differential}
\beq
dS = x \frac{dy}{y}\, , \ \oint_{\infty} dS = 0 
\eeq

Note the multi-valuedness of $y(x)$, a ubiquitous feature of the limit shapes: for finite ${\ve}_{1}, {\ve}_{2}$  the expectation value 
\eqref{eq:yvev} is a meromorphic function with poles and zeroes, but in the limit ${\ve}_{1}, {\ve}_{2} \to 0$ these condense, forming a cut. In the $A_1$-model this is simply a cut of a $2:1$ cover of the $x$-plane, described by \eqref{eq:swa1} \footnote{It is sometimes useful to represent \eqref{eq:swa1}
in the ancient Greek parametrization: 
\beq
\begin{aligned}
& y = (x-m_{+}) w\, , \ 
x = m_{+} \frac{w}{w-{\qe}} - m_{-} \frac{1}{w-1} \\
& dS = d \left( x + m_{+} \, {\rm log}(x-m_{+}) \right) + m_{-} \frac{dw}{w} + m_{+} \frac{dw}{w-{\qe}}
- m_{-} \frac{dw}{w-1}  
\end{aligned}
\label{eq:xwcurve}
\eeq}.

{}In the ${\hat A}_{0}$-case the structure of $y$-cuts is much more complicated. The so-called physical sheet where $y(x) \sim x + o(1)$, at $x \to \infty$, contains one cut. 
The analogue of \eqref{eq:swa1} reads as:
\begin{multline}
{\chi} (x):\, = \, y(x) \, {\phi}({\qe})\times \\
\sum_{{\nu} \in {\Lambda}}
\, {\qe}^{|{\nu}|}\, \prod_{{\square} \in {\nu}} \frac{y(x+c_{\square}(m, -m)-m)y(x+c_{\square}(m, -m)+m)}{y^2 (x+c_{\square}(m, -m))} \\
= x + {\hat u}_{0}
\label{eq:xa0vev}
\end{multline}
where we denoted ${\ve}_{3} = m$. 
The solution of this equation produces not just one, but two interesting analytic curves, the \emph{spectral curve} and the \emph{cameral curve} (cf. \cite{NP}). The way to resolve \eqref{eq:xa0vev} is to use the analogue of Jacobi identity \cite{GNncJ}: 
\begin{multline}
\sum_{M \in {\BZ}} (-1)^{M} z^{M} {\qe}^{\frac{M(M-1)}{2}} \, {\chi}(x- Mm)  = \\
y(x) \, {\phi}({\qe}) \, 
{\prod_{i=1}^\infty} \Big(1 - z^{-1} \cdot {\mathfrak{q}}^{i}  \frac{y_{i}(x)}{y_{i-1}(x)}   \Big) 
{\prod_{i=0}^\infty} \Big(1 - z\,  \cdot  {\mathfrak{q}}^{i}  \, \frac{y_{-i-1}(x)}{y_{-i}(x)}  \Big)  
\label{eq:thetatr}
\end{multline}
where we use the temporary short-hand notation $y_{i}(x) : = y(x+m{\cdot}i)$, $i \in {\BZ}$. Denoting the left hand side of
\eqref{eq:thetatr} by $R(z,x)$ we define the 
\emph{spectral curve} ${\CalC}^{\rm spec}$ as the zero locus 
\beq
R(z,x) = 0
\eeq
From the definition \eqref{eq:thetatr} it follows 
\beq
-z R(z{\qe}, x-m) = R(z,x)
\eeq
that ${\CalC}^{\rm spec}$ is ${\BZ}$-invariant: $(x,z) \mapsto (x-m, {\qe}z)$. 
Let ${\pi}_{x}, {\pi}_{z}$ denote the natural projections
\beq
{\pi}_{x}: {\CalC}^{\rm spec} \to {\BC}_{x}\, , \ {\pi}_{z}: {\CalC}^{\rm spec} \to {\BC}^{\times}_{z}\, , 
\eeq
e.g. for $p = (x,z) \in {\CalC}^{\rm spec}$, ${\pi}_{x}(p) = x, {\pi}_{z}(p)= z$. 
On the one hand, \eqref{eq:thetatr} gives a very explicit parametrization of ${\CalC}^{\rm spec}$ as a ${\BZ}$-cover of the elliptic curve ${\BC}^{\times}/{\qe}^{\BZ}$, $(x,z) \mapsto (z{\qe}^{\BZ})$. On the other hand, the right hand side of \eqref{eq:thetatr} implies that the fibers of ${\pi}_{x}$ are the branches ${\eig}_{n}(x)$, $n \in {\BZ}$, of the analytic continuation of 
\beq
{\eig}_{n}(x) \, = \, {\qe}^{n}\cdot {\eig}_{0}(x+m\cdot n) \, , \ {\eig}_{0}(x) = \frac{y(x)}{y(x-m)}
\eeq
The branch ${\eig}_{n}(x)$ approaches ${\qe}^{n}$
when $x \to \infty$. Also, for fixed $x$, 
${\eig}_{n}(x){\qe}^{-n} \to 1$ as $n \to \infty$. 
{}
In order to find the complete analytic continuation of $y(x)$, we need to study the 
\emph{cameral curve}. The theory of finite cameral covers has been developed in \cite{DonagiCam}. It will be partly reviewed in the section devoted to the $A_r$-models. In our problem, however, we need a more sophisticated notion, a \emph{semi-infinite spectral curve} ${\CalC}^{\rm spec}_{\frac{\infty}{2}}$:
\beq
{\CalC}^{\rm spec}_{\frac{\infty}{2}} \subset ( {\CalC}^{\rm spec} \times_{{\BC}_{x}} {\CalC}^{\rm spec} \times_{{\BC}_{x}} {\CalC}^{\rm spec} \times_{{\BC}_{x}} \ldots  )/S\left( {\infty} \right)
\label{eq:camc}
\eeq
We define it, set theoretically, as the set of equivalence classes of sequences ${\bf p} = (p_{n})$, $n=0, 1, \ldots$, 
$p_{n} \in {\CalC}^{\rm spec}$, such that: 
\beq
\begin{aligned}
{\sf i)} \qquad & {\pi}_{x}(p_{n}) = {\pi}_{x}(p_{0})\ {\rm for\ all}\ n >0 \, , \\
{\sf ii)} \qquad & {\pi}_{z}(p_{n}) {\qe}^{-n} \to 1\, , \ n \to \infty 
\end{aligned}
\label{eq:camcrv}
\eeq
modulo an equivalence relation: $(p_{n}) \sim (p'_{n})$ if there is a finite permutation 
${\sigma} \in S(N)$ for some $N \geq 0$, such that
\beq
\begin{aligned}
& p'_{n} = p_{n} \, , \ n \geq N \\
& p'_{n} = p_{{\sigma}(n)} \, , n < N \\
\end{aligned}
\label{eq:permutfin}
\eeq
We denote the group of such equivalences by $S\left( {\infty} \right)$. 
The analytic continuation of $y(x)$ is the function $Y$ on ${\CalC}^{\rm spec}_{\frac{\infty}{2}}$, defined as:
\beq
Y ({\bf p}) \, = \, {\rm Lim}_{N \to \infty} \, \left\lbrace \left( \prod_{n=0}^{N} {\pi}_{z}(p_{n}) \right) \, {\qe}^{-\frac{N(N+1)}{2}} \, \left( x - N {\ve}_{3} \right) \right\rbrace
\eeq
In the main body of the paper we shall give another, more constructive definition of the cameral curve and its semi-infinite quotient ${\CalC}^{\rm spec}_{\frac{\infty}{2}}$. 

$\bullet$ \emph{Higher times}

There is a natural generalization, studied in \cite{LMN, MN,GNVK},
\beq
{\mu}_{{\Lambda}; {\bf t}; {\hbar}} [{\lambda}] = e^{\sum_{k=1}^{\infty} t_{k}{\bf p}_{k}[{\lambda}]} \prod_{\square \in \lambda} \frac{({\Lambda} / {\hbar})^2}{h_{\square}^2} 
\label{eq:plancht}
\eeq
of the limit shape problem, with formal chemical potentials $(t_k)$, $k = 1, 2, \ldots$ for the generalized Casimirs: the observables ${\bf p}_{k}$ on $\Lambda$  given by the regularized powersums of $\lambda_{i} - i + \frac 12$, cf. \cite{OP}
\beq
{\bf p}_k [{\lambda}] = \frac{2^{-k}-1}{k+1} B_{k+1} + \sum_{i=1}^{{\ell}({\lambda})}  \left( {\lambda}_{i} - i  + {\scriptstyle{\frac 12}} \right)^{k} -  \left(  - i  + {\scriptstyle{\frac 12}} \right)^{k}\, .
\eeq
For example, 
\beq
{\bf p}_{1}[{\lambda}] = |{\lambda}| - \frac{1}{24}\, ,\ {\bf p}_{2}[{\lambda}] = \sum_{i=1}^{{\ell}({\lambda})}
{\lambda}_{i} ( {\lambda}_{i} + 1 - 2i ) \ .
\eeq

\subsection{The scope of this paper}
{}In this note we study several generalizations of \eqref{eq:macroplanch}, \eqref{eq:plancht} and their limit shapes. All of them are motivated by the studies of supersymmetric gauge theories in four dimensions \cite{Nekrasov:2002qd, NO}, and by topological string theory \cite{LMN}. Specifically, we study ${\hat A}_{r}$-models, and the $A_{r}$-models. The $A_r$-model can be obtained as a limit of ${\hat A}_{r+1}$ model, but it is methodologically useful to discuss it separately. 

These models are 
the ensembles of several partitions ${\lambda}^{({\si})}$, ${\si} = 0, \ldots, r$, interacting with each other in a cyclic/linear fashion, ${\lambda}^{({\si})} \leftrightarrow {\lambda}^{({\si}+1)}$, such that in the cyclic ${\hat A}_{r}$ case ${\lambda}^{(r+1)} = {\lambda}^{(0)}$, while in the linear $A_r$ case ${\lambda}^{(0)} = {\lambda}^{(r+1)} = {\emptyset}$. The dynamics is governed by the (complex, in general) probability measure ${\bmu}_{\bq; {\bt};  {\ba},  {\bve}} [ {\bla}]$, 
\beq
{\bve} = ({\ve}_{1},{\ve}_{2},{\ve}_{3},{\ve}_{4}) \, , \ {\ve}_{1}+{\ve}_{2}+{\ve}_{3}+{\ve}_{4} = 0\, , 
\label{eq:epss}
\eeq
which we introduce in the next section. 

The main goal of this paper is the analysis of the limit ${\ve}_{1},{\ve}_{2} \to 0$ shape and the emergent Whitham \cite{KricheverW} dynamics.

\subsection{Organization of the paper}

The paper is organized as follows. 

In section $\bf 2$ we introduce the ${\hat A}_{r}$-model, the measure ${\bmu}^{{\hat A}_{r}}_{{\bq}; {\bt} ; {\ba}, {\bve}} [ {\bla}]$. 
Through the suitable limit of the ${\hat A}_{r+1}$-model we define the $A_{r}$-model, 
the measure ${\bmu}^{A_{r}}_{{\bq}; {\bt} ; {\ba}, m_{\pm}, {\bve}} [ {\bla}]$. We recall the definition of $Y$-observables, $qq$-characters, and their limits, the $q$-characters and characters. Using the noncommutative Jacobi identity we proved in \cite{GNncJ} we organize the suitably (${\theta}$-)transformed characters into an infinite product in section $\bf 3$. We study this product in section $\bf 4$ in the limit ${\ve}_{1}, {\ve}_{2} \to 0$. We first do it on the small phase space, defined as the locus $\bt = 0$. The analysis brings us the so-called \emph{spectral curve}, encoding some information about the limit shape of the ensemble of partitions $\bla$. We find that in order to construct the complete analytic continuation of the
$Y$-observables $y_{\si}(x)$, ${\si} = 0, \ldots, r$, one needs to work with the \emph{cameral curve}. In the second part of section $\bf 4$ 
we turn back on the times $\bt$ and find the deformation of the limit shape using I.~Krichever's ideas. In section $\bf 5$ we briefly discuss the limit shape problem for the elliptic cohomology version of the ${\hat A}_{0}$-model on the small phase space. Section $\bf 6$ presents our conclusions.

\subsection{Acknowledgements}
 
 We have benefited from patient explanations of I.~Krichever (who is in fact the co-author of the section $\bf 4$) and A.~Okounkov. Research is partly supported by NSF PHY Award 2310279. 

\section{Measure and observables}

In this section, we recall the setup of the $\hat A_{r}$-model \cite{NP}. It is a rank one case of the gauge origami model \cite{BPSCFT3} associated with the orbifold group ${\BZ}_{r+1}$ acting on the ${\BC}^{2}_{34}$ space. For the uninitiated, we introduce it below. 

$\bullet$ \emph{The parameters}. Start by fixing $\bve$ as in \eqref{eq:epss}. 

{}The ${\hat A}_{r}$-model depends on $2(r+1)$ parameters $({\bq}, {\ba}) = ({\qe}_{\si}, a_{\si})_{{\si}=0}^{r}$, and
$r+1$ formal functions ${\bt} = ({\tau}_{\si}(x))_{{\si}=0}^{r}$, 
\beq
{\tau}_{\si}(x) = \sum_{k=0}^{\infty} {\tau}_{{\si}, k} \frac{x^{k}}{k!}\, , \ {\si} = 0, \ldots, r\, , 
\eeq
which we extend to all ${\si} \in {\BZ}$:
\beq
\begin{aligned}
& {\qe}_{{\si}+r+1} = {\qe}_{\si}\, , \ a_{{\si}+r+1} = a_{\si} - {\ve}_{3} \, , \\
& {\tau}_{{\si}+r+1}(x) = {\tau}_{\si}(x+{\ve}_{3}) \end{aligned}
\label{eq:aqt}
\eeq
The random variables are the $r+1$-periodic collections $\left({\lambda}^{({\si})}\right)_{{\si} \in {\BZ}}$, 
\beq
{\lambda}^{({\sf i}+r+1)} = {\lambda}^{({\sf i})}\, , 
\eeq
of partitions ${\lambda}^{({\si})} = \left( {\lambda}^{({\si})}_{i} \right)$.

$\bullet$ \emph{The ${\hat A}_{r}$-measure}.
{}Here is the expression for the measure:
\begin{multline}
{\bmu}_{\bq; \bt;  \ba, \bve}^{{\hat A}_{r}} [ {\bla}] \, = \,  \prod_{{\si}=0}^{r} \left( {\qe}_{\si}^{|{\lambda}^{({\si})}|} \, \prod_{{\square}=(i, j) \in {\lambda}^{({\si})}} e^{{\tau}_{\si}(a_{\si} + {\sf c}_{\square}({\ve}_{1}, {\ve}_{2}))} \right)\, \times \\
\prod_{{\si}=0}^{r}\prod_{{\square}=(i, j) \in {\lambda}^{({\si})}} 
\frac{\left( a_{\si}-a_{{\si}+1}+{\ve}_{1}({\lambda}_{j}^{({\si})t}+1-i) + {\ve}_{2}( j - {\lambda}_{i}^{({\si}+1)})\right)}{ \left( {\ve}_{1}({\sf l}_{\square}+1) - {\ve}_{2}{\sf a}_{\square} \right)}  \times \\
\prod_{{\si}=0}^{r}\prod_{{\square}=(i, j) \in {\lambda}^{({\si})}} 
\frac{\left( a_{{\si}-1}-a_{\si} + {\ve}_{1}(i - {\lambda}_{j}^{({\si})t}) + {\ve}_{2}({\lambda}_{i}^{({\si}-1)} + 1 - j) \right)}{ -{\ve}_{1}{\sf l}_{\square} + {\ve}_{2}({\sf a}_{\square}+1)}\ .
\label{eq:arhmeasure}
\end{multline}
We can view \eqref{eq:arhmeasure} as the statistical ensemble of a spin chain-type, 
with $r+1$ spins organized along a circle, interacting via nearest-neighbour interaction, with the 
spins being the partitions ${\lambda}^{({\si})}$. The reader might be familiar with 
the Heisenberg spin chain, based on the spin $\frac 12$ representations of 
$\mathfrak{sl}_{2}$. One can actually modify \eqref{eq:arhmeasure} by the additional
orbifold in the ${\BC}_{12}^{2}$ directions, replacing the sum over the infinite
set ${\Lambda}^{r+1}$ by that over a finite set of $2^{r+1}$ elements, making it into a version of spin $\frac 12$ spin chain with $r+1$ spins. We are not going to discuss this model in this paper though. 

Define  $z_{\si} \in {\BC}^{\times}$, ${\si} \in \BZ$ through:
\beq
{\qe}_{\si} = \frac{z_{{\si}+1}}{z_{\si}}\, , \ z_{{\si}+r+1} = {\qe} z_{\si}
\label{eq:zper}
\eeq
where
\beq
{\qe} = {\qe}_{0}{\qe}_{1} \ldots {\qe}_{r}
\label{eq:qutot}
\eeq

$\bullet$ \emph{Remarks}. There is some redundancy in the parametrization of \eqref{eq:arhmeasure}. First of all, 
the effective fugacity for the sizes $|{\lambda}^{({\si})}|$ of random partitions is
\beq
{\qe}_{\si} e^{{\tau}_{{\si},0}}
\eeq
There is however a difference in that we view ${\qe}_{\si}$ as complex numbers of absolute value less
than $1$, while ${\tau}_{{\si}, k}$, $k \geq 0$, are formal variables, i.e. we work over a ring
\beq
{\CalR} = {\BC}\left[ {\tau}_{{\si},k} \right]/{\tau}_{{\si},k}^{N_{{\si},k}}
\eeq
with sufficiently large $N_{{\si},k}$.

Secondly, the \emph{Coulomb parameters} $a_{\si} \in {\BC}$ enter the measure \eqref{eq:arhmeasure} in a way invariant under the overall shift
$(a_{\si}) \mapsto (a_{\si} + {\alpha})$, ${\alpha} \in {\BC}$, and the linear redefinition of the couplings ${\tau}_{{\si},k}$ associated with the argument shift ${\tau}_{\si}(x) \mapsto {\tau}_{\si}(x-{\alpha})$. Likewise, the parameters $z_{i} \in {\BC}^{\times}$ are defined up to an overall multiplicative shift $z_{i} \mapsto z_{i} {\lambda}$, $\lambda \in {\BC}^{\times}$. The space of gauge couplings of the ${\hat A}_{r}$-theory can be therefore identified with a coordinate patch in ${\CalM}_{1,r+1}$-the moduli space of $r+1$ points on a genus one curve ${\CalE}_{\qe} = {\BC}^{\times}/{\qe}^{\BZ}$, where the sequences $z_{{\si}+(r+1)p}$, $p \in {\BZ}$ 
define $r+1$ points $(z_{0}), (z_{1}), \ldots , (z_{r})$. This picture is in agreement with the general view on ${\CalN}=2$ dualities proposed in \cite{Gaiotto:2009we}.

$\bullet$ \emph{Times and times.} 
Introduce the power series in $x$, ${\sf t}_{\si}(x)$, ${\si} \in {\BZ}$ as solutions to the difference-functional equation:
\beq
{\sf t}_{{\si}{\sf -1}}(x) - 2{\sf t}_{\si}(x) + {\sf t}_{{\si}{\sf +1}}(x)  \, = \, {\rm log}({\qe}_{\si}) + {\tau}_{\si}(x) 
\label{eq:t34}
\eeq
together with the quasiperiodicity \eqref{eq:aqt}
\beq
{\sf t}_{{\si}{\sf +r+1}}(x) = {\sf t}_{\si}(x+{\ve}_{3}) \ . 
\eeq

$\bullet$ \emph{The $A_{r}$-measure}. Take the ${\hat A}_{r+1}$-model, and take the limit ${\qe}_{0}, {\qe}_{r+1} \to 0$. In this limit the only configurations with non-zero measure
have ${\lambda}^{(0)}= {\lambda}^{(r+1)} = {\emptyset}$. It is therefore meaningless to keep ${\tau}_{0}, {\tau}_{r+1}$, so we set these parameters to zero. Traditionally, we denote $a_{0} = m_{-}$, 
$a_{r+1} = m_{+}+{\ve}$, so that \eqref{eq:arhmeasure} becomes:
\begin{multline}
{\bmu}_{\bq; \bt;  \ba, m_{+}, m_{-}, \bve}^{A_{r}} [ {\bla}] \, = \, 
\prod_{{\si}=1}^{r} \left( {\qe}_{\si}^{|{\lambda}^{({\si})}|} \, \prod_{{\square}=(i, j) \in {\lambda}^{({\si})}} e^{{\tau}_{\si}(a_{\si} + {\sf c}_{\square}({\ve}_{1}, {\ve}_{2}))} \right)\, \times \\
\prod_{{\si}={\sf 1}}^{{\sf r-1}}\prod_{{\square}=(i, j) \in {\lambda}^{({\si})}} 
\frac{\left( a_{\si}-a_{{\si}+1}+{\ve}_{1}({\lambda}_{j}^{({\si})t}+1-i) + {\ve}_{2}( j - {\lambda}_{i}^{({\si}+1)}\right)}{ \left( {\ve}_{1}({\sf l}_{\square}+1) - {\ve}_{2}{\sf a}_{\square} \right)}  \times \\
\prod_{{\square}=(i, j) \in {\lambda}^{({\sf r})}} 
\frac{\left( a_{\sf r}+ {\sf c}_{\square}({\ve}_{1}, {\ve}_{2}) -m_{+} \right)}{ \left( {\ve}_{1}({\sf l}_{\square}+1) - {\ve}_{2}{\sf a}_{\square} \right)}  \times 
\prod_{{\square}=(i, j) \in {\lambda}^{({\sf 1})}} 
\frac{\left( m_{-}-a_{\sf 1} - {\sf c}_{\square}({\ve}_{1}, {\ve}_{2}) \right)}{ -{\ve}_{1}{\sf l}_{\square} + {\ve}_{2}({\sf a}_{\square}+1)}\\
\prod_{{\si}={\sf 2}}^{r}\prod_{{\square}=(i, j) \in {\lambda}^{({\si})}} 
\frac{\left( a_{{\si}-1}-a_{\si} + {\ve}_{1}(i - {\lambda}_{j}^{({\si})t}) + {\ve}_{2}({\lambda}_{i}^{({\si}-1)} + 1 - j) \right)}{ -{\ve}_{1}{\sf l}_{\square} + {\ve}_{2}({\sf a}_{\square}+1)}
\label{eq:armeasure}
\end{multline}

{}
$\bullet$ \emph{$Y$-observables for the $A_r$ and ${\hat A}_{r}$-models}. 
Using the $Y$-observables 
\eqref{eq:yobs2} we define:
\beq
Y_{\si}(x) [ {\bla}] := Y(x-a_{\si})[{\lambda}^{({\si})}]
\label{eq:yiobs}
\eeq
In the ${\hat A}_{r}$-case, thanks to \eqref{eq:aqt} we have the quasiperiodicity
\beq
Y_{{\si}+{\sf r+1}}(x) = Y_{\si}(x+{\ve}_{3})
\label{eq:yiarh}
\eeq
In other words, the $Y_{\si}(x)$-observable captures the shape of the $\si$'th partition
from the set $\bla$ of random partitions in the $\hat A_{r}$ or $A_{r}$-ensemble. 
We have the asymptotics:
\beq
Y_{\si}(x) = x - a_{\si} + \frac{\ve_{1}{\ve}_{2} k_{\si}}{x} + o(x^{-2})\, , \qquad x \to\infty
\label{eq:yxasympt}
\eeq
where $k_{\si}$ is the observable, the $\si$'th \emph{instanton charge}
\beq
k_{\si}[{\bla}] = |{\lambda}^{({\si})}|
\eeq

{}
$\bullet$ \emph{$qq$-characters: ${\hat A}_{r}$-model}
{}
The $\si$'th \emph{fundamental $qq$-character} of the ($\bt$-deformed) 
${\hat A}_{r}$-model is given by the ($\bt$-deformed) formula \cite{BPSCFT1, BPSCFT3}:
\begin{multline}
{\CalX}_{\si}^{{\hat A}_{r}} (x) =  
\, Y_{\si}(x+{\ve}) 
 \times \\
\sum_{\nu \in {\Lambda}} {\hat m}^{(r)}_{\bq, \bt ; {\ve}_{3}:{\ve}_{4}:{\ve}_{1}:{\ve}_{2}} [ {\nu}] \prod_{{\square} = (i,j) \in {\nu}} \frac{Y_{{\si}+i-j-1} \left( x+{\ve}(1-j) \right) }{Y_{{\si}+i-j}\left( x+{\ve}(1-j) \right)} \frac{Y_{{\si}+i-j+1} \left(x+{\ve}(2-j) \right)}{Y_{{\si}+i-j}\left(x+{\ve}(2-j) \right)}
\label{eq:xarh}
\end{multline}
where the \emph{unnormalized} measure is given by:
\begin{multline}
{\hat m}^{(r)}_{\bq, \bt ; {\ve}_{3}:{\ve}_{4}:{\ve}_{1}:{\ve}_{2}} [ {\nu}]
\, = \,
\prod_{{\square} = (i,j) \in {\nu}} {\qe}_{{\si}+i-j} e^{{\tau}_{{\si}+i-j}(x+{\ve}(1-j))}\, \times \\ \\
\prod_{{\square}\in {\nu}, \, {\sf h}_{\square} \equiv 0 (r+1)} 
\left( 1 + \frac{\ve_1 \ve_2}{\left({\tilde\ve}_{3}{\sf h}_{\square} + {\ve}{\sf a}_{\square}\right) \left( {\tilde\ve}_{3}{\sf h}_{\square} + {\ve} ({\sf a}_{\square}+1) \right)} \right)
\end{multline}
with
\beq
{\tilde\ve}_{3} = \frac{\ve_3}{r+1}
\eeq
The main theorem \cite{BPSCFT2} about \eqref{eq:xarh} states that the expectation values of the $qq$-characters have no poles in $x$. In the limit ${\ve}_{1}, {\ve}_{2} \to 0$ this becomes a system of analytic (algebraic in power series expansion in ${\qe}_{\si}$'s) relations between the multi-valued analytic functions 
\beq
y_{\si} (x) = \langle Y_{\si}(x) \rangle_{\bq; \bt;  \ba, \bve}^{{\hat A}_{r}} \, , \ 
\eeq
and the \emph{characters} \cite{NP}
\begin{multline}
{\chi}_{\si}(x) = \, \frac{\left\langle {\CalX}_{\si}^{{\hat A}_{r}} (x) \right\rangle_{\bq; \bt;  \ba, \bve}^{{\hat A}_{r}}}{\langle 1 \rangle_{\bq; \bt;  \ba, \bve}^{{\hat A}_{r}}} \biggr\vert_{{\ve}_{1},{\ve}_{2}\to 0} \\
= y_{\si}(x) \sum_{\nu \in {\Lambda}}\, \prod_{{\square} = (i,j) \in {\nu}} {\qe}_{{\si}+i-j} e^{{\tau}_{{\si}+i-j}(x)}\,  \frac{y_{{\si}+i-j-1} \left( x\right) }{y_{{\si}+i-j}\left( x \right)} \frac{y_{{\si}+i-j+1} \left(x \right)}{y_{{\si}+i-j}\left(x \right)}\, , 
\label{eq:affchar}
\end{multline}
 which are the entire functions of $x$. For future use, define
 the \emph{generalized eigenvalues}
\beq
{\eig}_{\si} (x) : = z_{\si}  e^{{\xi}_{\si}(x)} \frac{y_{\si}(x)}{y_{{\si}{\sf -1}}(x)} \, , \ {\si} \in {\BZ}
\label{eq:geneig}
\eeq
obeying the twisted periodicity
\beq
{\eig}_{{\si} + r+1} (x) = {\qe} {\eig}_{\si}(x+{\ve}_{3})\, , 
\eeq
where the formal\footnote{The word ``formal'' here means a power series in $x$ whose coefficients are linear in the formal variables ${\tau}_{{\si}, k}$. } functions ${\xi}_{i}(x)$, $i \in {\BZ}$, are related to ${\sf t}_{\si}(x)$ via
\beq
{\xi}_{\si}(x) ={\sf t}_{\si}(x) - {\sf t}_{{\si}{\sf -1}}(x) - {\rm log}({\qe}) \frac{x}{\ve_3} - {\rm log}(z_{\si})
\label{eq:sixapprox}
\eeq
As a consequence, the $\xi_{\si}$'s
solve the first-order difference equations
\beq
{\xi}_{{\si}{\sf +1}}(x) - {\xi}_{\si}(x) = {\tau}_{\si}(x)\, , \ \ . 
\eeq
and obey the familiar quasiperiodicity
\beq
{\xi}_{i+r+1}(x) = {\xi}_{i}(x+{\ve}_{3}) \ .
\eeq
{}
$\bullet$ \emph{$qq$-characters: $A_{r}$-model}
{}
The $\si$'th \emph{fundamental $qq$-character} of the ($\bt$-deformed) 
$A_{r}$-model is given by the ($\bt$-deformed) formula \cite{BPSCFT1, BPSCFT3}, which is obtained 
from \eqref{eq:xarh} for the ${\hat A}_{r+1}$-model by specifying ${\qe}_{0} = {\qe}_{r+1} = 0$: 
for ${\si} = 1, \ldots, r$: 
\begin{multline}
{\CalX}_{\si}^{A_{r}} (x) =  
\, Y_{\si}(x+{\ve}) 
\sum_{\nu \in {\Lambda}_{{\si},r}} \prod_{{\square} = (i,j) \in {\nu}} {\qe}_{{\si}+i-j} e^{{\tau}_{{\si}+i-j}(x+{\ve}(1-j))}\,  \\ \times \, \prod_{{\square} = (i,j) \in {\nu}} \frac{Y_{{\si}+i-j-1} \left( x+{\ve}(1-j) \right) }{Y_{{\si}+i-j}\left( x+{\ve}(1-j) \right)} \frac{Y_{{\si}+i-j+1} \left(x+{\ve}(2-j) \right)}{Y_{{\si}+i-j}\left(x+{\ve}(2-j) \right)}
\label{eq:xar}
\end{multline}
where
${\Lambda}_{{\si},r}$ is a finite set of partitions $\nu$, such that for every $\square = (i,j) \in \nu$, $0 < {\si} + i - j \leq r$, equivalently the Young diagram of $\nu$ fits a ${\si}(r+1-{\si})$
rectangle: 
\beq
{\nu}_{1}^{t} \leq r+1 - {\si} \, , \ {\nu}_{1}  \leq {\si}
\eeq
There is a bijection between ${\Lambda}_{{\si}, r}$ and the set of all cardinality $\si$ subsets of the set
$\{ 1, \ldots, r+1 \}$: 
\beq
{\Lambda}_{{\si}, r} = \left\{ \, I \, | \, |I| = {\si} \, , \ I = \{ {\iota}_{1}, \ldots , {\iota}_{\si} \} \subset \{ 1, \ldots , r+1 \} \, \right\}
\label{eq:lir}
\eeq
\beq
{\iota}_{j} = {\nu}_{j}^{t} - j +{\si}+1\, , \ r+1 \geq {\iota}_{1} > {\iota}_{2} > \ldots > {\iota}_{\si} \geq 1
\eeq
The limit ${\qe}_{0}, {\qe}_{r+1} \to 0$ suggests to keep $z_{1}, \ldots, z_{r+1} \in {\BC}^{\times}$ finite, with
\beq
z_{i} = 0\, , \ i > r+1\, ,\ z_{i} = {\infty}\, , \ i < 0
\label{eq:zcoordar}
\eeq
parametrizing a coordinate patch in ${\CalM}_{0,r+3}$, in agreement with \cite{Gaiotto:2009we}. 
In the ${\ve}_{1}, {\ve}_{2}\to 0$ limit the $\si$'th character of the $A_r$-model simplifies, 
\beq
{\chi}_{\si}(x) \, = \, \frac{\left\langle {\CalX}_{\si}^{A_{r}} (x) \right\rangle_{\bq; \bt;  \ba, m_{\pm}, \bve}^{A_{r}}}{\langle 1 \rangle_{\bq; \bt;  \ba, m_{\pm}, \bve}^{A_{r}}} \biggr\vert_{{\ve}_{1},{\ve}_{2}\to 0}
\eeq
so that
\beq
{\prod\limits_{{\sj}=1}^{\si} z_{\sj}e^{{\xi}_{\sj}(x)}}\, {\chi}_{\si}(x) \, = \, y_{0}(x) \, e_{\si}
\left( {\eig}_{1}(x), \ldots, {\eig}_{r+1}(x) \right)  
\label{eq:chies}
\eeq
is $y_{0}(x) = x - m_{-}$ times 
the $\si$'th elementary symmetric function of ${\eig}_{\sf 1}, \ldots, {\eig}_{\sf r+1}$, defined in 
 
 \section{Product formulas}
 
 To solve the inverse problem, i.e. recovering $y_{\si}$'s from $\chi_{\si}$'s define the ${\BZ} \times {\BZ}$ matrix $\bf Y$, cf. \eqref{eq:t34}:
 \beq
 {\bf Y}_{i,j} = y_{i-j}(x) \, e^{{\sf t}_{i-j}(x)}\, , \ i,j \in {\BZ}\, , 
 \label{eq:ymat}
 \eeq

The significance of \eqref{eq:ymat} is  \cite{GNncJ}:
\beq
{\chi}_{\si}(x)\, e^{{\sf t}_{\si}(x)}\ = \ \sum_{\nu \in {\Lambda}} {\CalX}_{\nu} \left[ ^{({\si})}{\bf Y} \right] 
\label{eq:charfrommat}
\eeq
where
\beq
^{({\si})}{\bf Y}_{a,b} = {\bf Y}_{a+{\si},b}
\eeq
\subsection{Affine ${\hat A}_{r}$-case}
Now we can use the generalized Jacobi identity \cite{GNncJ} and the notation \eqref{eq:geneig} to prove the identity
relating two convergent power series in $\qe$ over $\CalR$:
\begin{multline}
\sum_{M \in {\BZ}} \,  {\chi}_{M}(x)\, (-z)^{-M}\, {\CalQ}_{M} \, e^{{\Xi}_{M}(x)}
\, = \, \\
= (-z)^{-\si} y_{\si}(x) \, \prod_{l > {\si}} \left(1 - \frac{{\eig}_{l}(x)}{z} \right) \, \prod_{l \leq {\si}} \left(1 - \frac{z}{{\eig}_{l}(x)}  \right) \,  ,
\label{eq:ncj}
\end{multline}
for any ${\si} \in {\BZ}$. 
We call the left hand side of \eqref{eq:ncj} the \emph{$\theta$-transform of affine characters ${\chi}_{\si}$}. Let us unpack \eqref{eq:ncj}: there is an indeterminate
$z$, we defined the $z_i$'s in \eqref{eq:zper},
and ${\CalQ}_{M}$, ${\Xi}_{M}(x)$ are given by (cf. \eqref{eq:sixapprox})\footnote{In convenient notation of A.~Givental, 
\beq
{\CalQ}_{M} = \frac{\prod\limits_{-\infty}^{M} z_{i}}{\prod\limits_{-\infty}^{0} z_{i}} \, ,\qquad
{\Xi}_{M}(x) = \sum_{-\infty}^{M} {\xi}_{i}(x)  - \sum_{-\infty}^{0} {\xi}_{i}(x)
\eeq}:
\beq
{\Xi}_{M}(x) = \begin{cases} \,   -\sum\limits_{i=M+1}^{0} {\xi}_{i}(x)   & \, , \qquad M < 0 \\
\qquad  0 & \, , \qquad M = 0\\
\qquad  \sum\limits_{i=1}^{M} {\xi}_{i}(x)  & \, , \qquad M > 0 \end{cases}\, , \qquad
{\CalQ}_{M} = \begin{cases} \,   \prod\limits_{i=M+1}^{0} z_{i}^{-1}   & \, , \qquad M < 0 \\
\qquad  1 & \, , \qquad M = 0\\
\quad  \prod\limits_{i=1}^{M} z_{i}  & \, , \qquad M > 0 \end{cases} \ , 
\label{eq:qxifuns}
\eeq
so that
\beq
{\CalQ}_{M} \, e^{{\Xi}_{M}(x)} \, =\,  
e^{{\sf t}_{M}(x)- {\sf t}_{0}(x)} \, {\qe}^{-\frac{Mx}{\ve_3}} \ .
\eeq
Using \eqref{eq:zper}, we can simplify, for
$\si = 0, \ldots, r$, $p \in {\BZ}$, 
\beq
{\CalQ}_{{\si} + p(r+1)} = \left( \prod_{{\sj} =1}^{\si} z_{\sj} \right) \, \left( z_{*}^{r+1} \right)^{p} \, {\qe}^{\frac{p(p+1)}{2}}\, , \
\eeq
where the ``center-of-mass'' is defined via
\beq
z_{*}^{r+1} = \prod_{{\sj} = 1}^{r+1} z_{\sj} \ . 
\eeq
For example
\beq
{\CalQ}_{0} = 1\, , \ {\CalQ}_{\sf 1} = z_{\sf 1}\, , \ {\CalQ}_{\sf 2} = z_{\sf 1}z_{\sf 2}\, , \ \ldots \, , \ {\CalQ}_{\sf r} = z_{*}^{r+1} z_{\sf r+1}^{-1}
\eeq
\subsection{Finite $A_r$-case}

Using \eqref{eq:chies} the identity \eqref{eq:ncj} becomes the Vieta's theorem for the \emph{spectral polynomial}
\beq
y_{0} (x) \prod_{{\sa}=1}^{r+1} \left( 1 - \frac{{\eig}_{\sa}(x)}{z} \right) = \sum_{M=0}^{r+1} (-z)^{-M} {\chi}_{M}(x) \prod_{{\sa}=1}^{M} z_{\sa} e^{{\psi}_{\sa}(x)}
\label{eq:speccurvf}
\eeq
with ${\chi}_{0}(x) = y_{0}(x)$, ${\chi}_{r+1} (x) = y_{r+1}(x) = x - m_{+}$. 

\section{The Limit Shape}

We shall now address the question of the limit shape. From \eqref{eq:yxasympt} we know the asymptotics of
$y_{\si}(x)$'s on the physical sheet (we shall explain below in more detail, what are the sheets of the analytic continuations of $\langle Y_{\si}(x) \rangle$'s):
\beq
y_{\si}(x) = x - a_{\si} + o(1) \, , \  x \to \infty
\label{eq:yiasym}
\eeq
This asymptotics implies the following expansion of ${\eig}_{i}(x)$ near $x = {\infty}$: in the ${\hat A}_{r}$-case:
\beq
{\eig}_{{\si} + p(r+1)}(x) \to z_{\si} {\qe}^{p} \left( 1 + \frac{a_{{\si}{\sf -1}}-a_{\si}}{x} + \ldots \right) \cdot {\exp} \, \sum_{k=0}^{\infty} {\xi}_{{\si},k} (x+{\ve}_{3}p)^{k} \, , \qquad x \to \infty 
\label{eq:eigxas}
\eeq
with $p \in {\BZ}$, e.g. ${\si} = 1, \ldots, r+1$, 
and $\ldots$ stand for $x^{-1}$ times a power series in $x^{-1}$, and 
in the $A_r$-case:
\beq
{\eig}_{{\si}}(x) \to z_{\si}  \left( 1 + \frac{a_{{\si}{\sf -1}}-a_{\si}}{x} + \ldots \right) \cdot {\exp} \, \sum_{k=0}^{\infty} {\xi}_{{\si},k} x^{k} \, , \qquad x \to \infty  \ .
\label{eq:eigxasf}
\eeq

\subsection{Small phase space}

We start with the \emph{small phase space} problem, ${\tau}_{\si}(x) = 0$, ${\xi}_{\si}(x) = 0$, in which case the problem of expressing $y_{\si}$'s through ${\chi}_{\si}$'s is simple enough. 
First of all, the absence of poles and the large $x$ asymptotics \eqref{eq:yxasympt} implies
that
\beq
{\chi}_{\si}(x) = {\vt}_{\si} ({\bq}) \left( x + \frac{\si}{r+1}{\ve}_{3} + {\hat u}_{\si}({\bq}; {\ba}) \right)\ , 
\eeq
with  
\beq
{\vt}_{\si}({\bq}) = \sum_{\nu \in \Lambda} \, \prod_{j=1}^{{\nu}_{1}^{t}} \frac{z_{{\si}+{\nu}_{j}^{t}-j+1}}{z_{{\si}-j+1}} = {\vt}_{{\si}+r+1}({\bq})
\eeq
in the ${\hat A}_{r}$-case, and
\beq
{\chi}_{\sa}(x) = \frac{e_{\sa}(z_{1}, \ldots, z_{r+1})}{z_{1}\ldots z_{\sa}} \left(x+u_{\sa}({\bq}; {\ba}) \right)\, , \ {\sa} = 1, \ldots , r+1
\label{eq:symma}
\eeq
in the $A_r$-case. Here ${\hat u}_{\si}$, $u_{\si}$  are some constants (in $x$) obeying ${\hat u}_{{\si}+r+1} = {\hat u}_{\si}$ in the affine ${\hat A}_{r}$-case, and
$u_{0} = -m_{-}$, $u_{r+1} = - m_{+}$ in the $A_r$-case. 
Then, the $\theta$-transform of affine characters can be computed explicitly, giving the identity:
\beq
{\sf P}(z, x)  \, = \, 
{\Theta}(z; {\bq}) \, {\sf R}(z, x) 
\label{eq:ncj3}
\eeq
for the product
\beq
{\sf P}(z, x) : = y_{0}(x) \, {\prod_{i=1}^\infty} \left(1 - \frac{{\eig}_{i}(x)}{z} \right) 
{\prod_{i=0}^\infty} \left(1 - \frac{z}{{\eig}_{-i}(x)} \right)  \\
\label{eq:affprod}
\eeq
where
\beq
{\Theta}(z; {\bq}) = \prod_{i=1}^{\infty} \left(1 - \frac{z_{i}}{z} \right) \left(1 - \frac{z}{z_{1-i}}  \right) \,
\label{eq:rtheta}
\eeq
can be expressed in terms of the theta function \eqref{eq:oddtheta}, 
and\footnote{We use that every integer $i \in {\BZ}$ can be uniquely represented as
\beq
i = {\si} + p (r+1)\, , \ {\si} = 1, \ldots, r+1\,  , \ p \in {\BZ} 
\eeq
where $p \geq 0$ for $i > 0$, and $p < 0$ for $i \leq 0$.}
\beq
{\sf R}(z, x)\, = \, 
 x - {\sf a} + \sum_{{\si}=1}^{r+1} (a_{\si} - a_{{\si}{\sf -1}}) {\zeta}(z/z_{\si}) 
 \label{eq:rzx}
\eeq
where 
\beq
{\sf a} = \frac{a_{\sf r+1} + a_{\sf 0}}{2} = a_{\sf 0} - \frac{\ve_3}{2}
\eeq
and
\begin{multline}
{\zeta}(z) = \frac 12 + \sum_{p =1}^{\infty} \left( \frac{{\qe}^{p-1}}{z- {\qe}^{p-1}}  +  \frac{z}{z-{\qe}^{-p}} \right) \, , \\
{\zeta}(z)= - {\zeta}(z^{-1})  \, , \ 
{\zeta}({\qe}z) = {\zeta}(z) -1
\label{eq:zeta}
\end{multline}
It is obvious from the definition \eqref{eq:rtheta}
that
\beq
{\Theta}({\qe}z; {\bq}) = \left(-\frac{z_{*}}{{\qe} z} \right)^{r+1} \, {\Theta}(z; {\bq})\, , \ 
\label{eq:rtheta2}
\eeq
while from \eqref{eq:zeta} we deduce the invariance of ${\sf R}$ under the $\BZ$-action
\beq
{\sf R}(x-{\ve}_{3}, {\qe}z) = {\sf R}(x,z)
\eeq
Thus, ${\Theta}_{\bq} := {\Theta}(z; {\bq})$ is a holomorphic section of degree $r+1$ line bundle ${\CalL}_{\bq}$ over ${\CalE}_{\qe}$, vanishing at the $r+1$ points $(z_{\sf 0}), (z_{\sf 1}), \ldots, (z_{\sf r})$. Such section is unique up to a multiplicative constant. The set of gauge couplings of
the  ${\hat A}_{r}$-theory is therefore identified with the space of pairs $\left( {\CalE}_{\qe}, [{\Theta}_{\bq}] \in {\BP}H^{0}({\CalL}_{\bq}) \right)$. 

$\bullet$ \emph{Remark}. 
From now on we shall use the redundancy in definition of $z_{i}$'s to set 
\beq
z_{*} = 1
\label{eq:cm}
\eeq

\subsection{Spectral curve}

We thus have the following {\bf Theorem}: for the ${\hat A}_{r}$-model the analytic continuations of ${\eig}_{i}(x)$, $i \in \BZ$ defined by \eqref{eq:geneig}, are the branches of the \emph{spectral curve}
\beq
x = {\sf a} + \sum_{{\si}=1}^{\sf r+1} (a_{\sf i-1} - a_{\si}) {\zeta}(z/z_{\si}) \ .
\label{eq:speccahr}
\eeq
The moments of the limit shape, i.e. the coefficients $ch_{{\si},k}$
of the large $x$ expansion 
\beq
y_{\si}(x) = x  \, {\exp} \, - \sum_{k=1}^{\infty} \frac{k!}{kx^{k}} ch_{{\si},k}
\eeq
can be extracted from the periods
\beq
\frac{1}{2\pi\ii}\oint_{C_{\si}} x^{k} \frac{dz}{z}
\eeq
around the small circles $C_{\si} = \{ \, z \, | \, |z-z_{\si}| = {\delta} \downarrow 0 \, \}$. 

To get to the $A_r$-case we take the $\qe \to 0$ limit of the \eqref{eq:speccahr} with $r$ replaced by $r+1$, while keeping $z_{1}, \ldots, z_{r+1}$ finite. Alternatively, we
can directly compute both sides of the identity \eqref{eq:speccurvf} with $\psi_{\ba} = 0$. Either way we get:
\beq
x = a_{\sf 0} + \sum_{{\si}=1}^{\sf r+1}  \frac{z_{\si}(a_{\sf i-1} - a_{\si}) }{z-z_{\si}}
\label{eq:speccar}
\eeq
The differential 
\beq
dS := x \frac{dz}{z} = dz \sum_{{\ba}= 0}^{r+2} \frac{p_{\ba}}{z-z_{\ba}}
\eeq
with $z_{0} = \infty$, $z_{r+2}  = 0$, $p_{r+2} = a_{r+1}$, 
$p_{\si} = a_{\sf i-1} - a_{\si}$, 
\beq
\sum_{\ba = 0} p_{\ba} = 0
\eeq
encodes the Coulomb parameters as periods around $z_{\si}$'s, and $ch_{{\si},k}$'s are again computed from the periods of $x^{k-1} dS$. 

\subsection{Cameral curve(s)}

Despite the fact that the spectral curve implicitly contains all the information about the limit shape, it might be useful to have a more specific characterization of the analytic continuation of $y_{\si}(x)$. 

It is instructive to start the discussion in the $A_r$-case. The curve ${\CalC}^{\rm spec}$ defined by the Eq. \eqref{eq:speccar} is rational, with the map ${\pi}_{x}$ being $r+1 : 1$, as can be easily seen by converting the Eq. \eqref{eq:speccar} into a degree $r+1$ polynomial 
\beq
{\sf P}(z, x) = \prod_{{\si} =1}^{r+1} (z- z_{\si} ) \left( x - a_{\sf 0} + \sum_{{\si} =1}^{r+1} 
   \frac{z_{\si}(a_{\si} - a_{\sf i -1})}{z-z_{\si}} \right)
\label{eq:specrvar}
\eeq
in $z$ with coefficients linear in $x$.  The Eq. \eqref{eq:symma} can be interpreted as equations for ${\eig}_{\ba} = {\eig}_{\ba}(x)$:
\beq
e_{\ba}\left( {\eig}_{\sf 1}, \ldots  , {\eig}_{\sf r+1} \right) = e_{\ba}\left( z_{\sf 1}, \ldots  , z_{\sf r+1} \right) \, \frac{x + u_{\ba}}{x-m_{-}}\, , \ {\ba} = 1, \ldots, r+1
\label{eq:symmb}
\eeq
Define the \emph{cameral} curve ${\CalC}^{\rm cam}$ as the curve in $\left( {\BC\BP}^{1}_{\eig} \right)^{r+1} \times {\BC\BP}^{1}_{x}$
defined by $r+1$ equations \eqref{eq:symmb}. By design, this curve is invariant under the action of the symmetric group $S(r+1)$, permuting ${\eig}_{\ba}$'s. We can define many curves by taking the quotients of ${\CalC}^{\rm cam}$ by subgroups of $S(r+1)$. For example, the spectral curve ${\CalC}^{\rm spec}$ is a quotient of ${\CalC}^{\rm cam}$ by the action of $S(r)$ permuting ${\eig}_{\sf 2}, \ldots , {\eig}_{\sf r+1}$. Knowing $z = {\eig}_{\sf 1}$ we can reconstruct the (analytic continuation of) $y_{\sf 1}(x)$, by
\beq
y_{\sf 1}(x) = y_{\sf 0}(x) z
\eeq
Define the $\si$'th fundamental spectral curve ${\CalC}^{\rm spec}_{\si}$
with ${\CalC}^{\rm spec}_{\sf 1} = {\CalC}^{\rm spec}$ as a quotient of ${\CalC}^{\rm cam}$
by the action of $S({\si}) \times S({\sf r+1-i})$ permuting $\left( {\eig}_{\sf 1}, \ldots, {\eig}_{\sf i}\right)$, and $\left( {\eig}_{\sf i+1}, \ldots {\eig}_{\sf r+1} \right)$
separately\footnote{The $\si$'th spectral curve encodes $y_{\si}(x)$ via
\beq
y_{\si}(x) = y_{\sf 0}(x) {\sigma}_{\si}^{-1}
\label{eq:yixfrom}
\eeq}. 
For example, for ${\si}=1$, we would get
\beq
(1-{\sigma}_{1}z) y_{\sf 1} z^{r}  R_{r}(z^{-1}) = - {\sf P}(z,x)
\eeq
A way to solve this equation is to note that $u = {\sigma}_{1}^{-1}$ is a root of ${\sf P}(z,x)$ in $z$, ${\sf P}(u,x) = 0$. Given such $u$, $y_{\sf 1}$ can be computed as 
\beq
y_{\sf 1}(u,x) = - \frac{1}{2\pi\ii} 
\oint_{C_{\infty}} \frac{dz}{z^{r+1}} \frac{{\sf P}(z,x)}{1-z/u} = y_{0}(x) u
\eeq
In other words, the Riemann surface of $y_{\sf 1}(x)$ is the algebraic curve
\beq
{\sf P}\left( \frac{y_{\sf 1}}{x - m_{-}} , x \right) = 0
\eeq
which we already know.
{}
For ${\si}>1$, we need a more general construction. In fact, we present two constructions, one is explicitly algebraic, another is more analytic in nature (and has the advantage of being generalizable to the affine case). 

\begin{enumerate}

\item
Given a polynomial ${\sf P}(z,x)$ in $z$ and $x$, of degrees $r+1$ and $N$, respectively, 
define\footnote{For example, 
 define
three polynomials ${\sf P}^{(3)}_{a}$ , $a = 1, 2,3$ in four variables ${\tau}_{1},{\tau}_{2},{\tau}_{3}, x$:
\beq
\begin{aligned}
& {\sf P}_{1}^{(3)}( {\boldsymbol{\tau}}; x) = {\sf P} \left(z_{1} ,x \right) + {\sf P}\left(z_{2} ,x \right)+ {\sf P}\left(z_{3} ,x \right)\, , \\
&  {\sf P}_{2}^{(3)} ( {\boldsymbol{\tau}}; x) = \frac{{\sf P} \left(z_{1} ,x \right) - {\sf P}\left(z_{2} ,x \right)}{z_{1}-z_{2}} + \frac{{\sf P} \left(z_{1} ,x \right) - {\sf P}\left(z_{3} ,x \right)}{z_{1}-z_{3}} + \frac{{\sf P} \left(z_{2} ,x \right) - {\sf P}\left(z_{3} ,x \right)}{z_{2}-z_{3}}\, , \\
&  {\sf P}_{3}^{(3)} ( {\boldsymbol{\tau}}; x) = \frac{\frac{{\sf P} \left(z_{1} ,x \right) - {\sf P}\left(z_{2} ,x \right)}{z_{1}-z_{2}} - \frac{{\sf P} \left(z_{1} ,x \right) - {\sf P}\left(z_{3} ,x \right)}{z_{1}-z_{3}}}{z_{2}-z_{3}} = 
\frac{{\sf P}(z_{1},x)}{z_{12}z_{13}} + \frac{{\sf P}(z_{2},x)}{z_{21}z_{23}} + \frac{{\sf P}(z_{3},x)}{z_{31}z_{32}} \, , \\
& {\tau}_{1} = z_{1}+z_{2}+z_{3}\, , \ {\tau}_{2} = z_{1}z_{2} + z_{1}z_{3} +z_{2}z_{3}\,  ,\ {\tau}_{3} = z_{1}z_{2}z_{3} \, , 
\end{aligned}
\label{eq:P3}
\eeq}
${\si}$ polynomials ${\sf P}^{({\si})}_{a}$, $a = 1, \ldots, {\si}$, in ${\si}+1$ variables $({\boldsymbol{\tau}}, x) = ({\tau}_{1}, \ldots , {\tau}_{\si}, x)$:
\beq
\begin{aligned}
& {\sf P}^{({\si})}_{a}(\boldsymbol{\tau}, x) = \\
& \qquad\qquad = \sum_{k=1}^{\si} \frac{e_{{\si}-a}(z_{1,k}, \ldots,
z_{k-1,k}, z_{k+1,k}, \ldots, z_{{\si},k})}{\prod\limits_{j\neq k} \, ( z_{k}-z_{j} )} {\sf P}(z_{k}, x) \\
& \qquad z_{n,m} = z_{n} -z_{m} \, , \\
& \sum_{a=0}^{\si} w^{a} {\sf P}^{({\si})}_{a}(\boldsymbol{\tau}, x) \,  = \, \frac{1}{2\pi\ii}  \oint_{{\rm around\ zeroes\ of}\ T(z) } \, dz\, {\sf P}(z,x)\, \frac{T(z+w) }{T(z)} \\  
& T(z) := \prod_{l=1}^{\si} (z-z_{l}) =  z^{\si} - {\tau}_{1} z^{{\si}-1} + {\tau}_{2}z^{{\si}-2} + \ldots + (-1)^{\si} {\tau}_{\si} \, , \\
& \qquad\qquad {\tau}_{m} = e_{m}(z_{1}, \ldots, z_{\si}) \, ,\ m = 1, \ldots, {\si} 
\end{aligned}
\label{eq:symmpc}
\eeq
of multi-degrees $(r+2-a, \left[ \frac{r+2-a}{2} \right], \ldots, \left[ \frac{r+2-a}{\si} \right], N )$ in $({\boldsymbol{\tau}}, x)$. 

The $\si$'th fundamental spectral curve is given by $\si$ polynomial equations in ${\si}+1$ variable:
\beq
P_{a}^{({\si})}({\boldsymbol{\tau}}, x) = 0 \, , \ a = 1, \ldots, {\si}
\eeq
The analytic continuation of $y_{\si}(x)$ is the function $y_{\si}$ on ${\CalC}^{\rm spec}_{\si}$
given by
\beq
y_{\si} = {\tau}_{\si} \, y_{0}(x)\, , 
\label{eq:ysitausi}
\eeq
\item
Let us reformulate the above construction.
Given $r+1$ variables ${\eig}_{1}, \ldots, {\eig}_{r+1} \in {\BC}^{\times}$ introduce the $S({\si}) \times S({\sf r+1-i})$-invariants:

{}
${\sigma}_{\sf 1}, {\sigma}_{\sf 2}, \ldots, {\sigma}_{\sf r+1-i}$ are the elementary symmetric polymomials of ${\eig}_{\sf 1}^{-1}, \ldots, {\eig}_{\sf r+1-i}^{-1}$, and 

{}
${\tau}_{\sf 1}, \ldots, {\tau}_{\si}$ are the elementary symmetric polynomials of ${\eig}_{\sf r+2-i}, \ldots, {\eig}_{\sf r+1}$. The curve ${\CalC}^{\rm spec}_{\si}$ is described by $r+2$ equations
on $r+2$ variables 
\[ \left( y_{\si}; {\sigma}_{\sf 1}, \ldots, {\sigma}_{\sf r+1-i}; {\tau}_{\sf 1}, \ldots, {\tau}_{\si}\right) \] 
over ${\BC\BP}^{1}_{x}$, obtained by equating the coefficients of $z^{-l}$, $l = 0, \ldots, r+1$ on both sides of the equation below:
\begin{multline}
{\Phi}_{+}(z; {\boldsymbol{\sigma}}) \cdot y_{\si} \cdot {\Phi}_{-}(z; {\boldsymbol{\tau}}) \,
= \, (-z)^{-{\si}} {\sf P}(z, x) \, , \\
{\Phi}_{+}(z; {\boldsymbol{\sigma}}) = 1 - z  {\sigma}_{1} + \ldots  + (-z)^{\sf r+1-i} {\sigma}_{\sf r+1-i}   \, , \\
{\Phi}_{-}(z; {\boldsymbol{\tau}}) \, =  \, 1 - z^{-1} {\tau}_{\sf 1}  + \ldots
+ (-z)^{-{\si}} {\tau}_{\si}  \\
\label{eq:factor1}
\end{multline}
{}
Now let us recast the factorization problem \eqref{eq:factor1} in a geometric way. 
Let $\gamma$ be a closed contour on ${\BC}^{\times}_{z}$ separating $z= 0$ and the zeroes of ${\Phi}_{-}$ from $z = {\infty}$ and  the zeroes of ${\Phi}_{+}$.  Then it is easy to see that the arguments of ${\Phi}_{+}$ and ${\Phi}_{-}$ are single-valued on $\gamma$,  
\beq
\oint_{\gamma} \, d{\rm log}\,{\Phi}_{+} = 
\oint_{\gamma} \, d{\rm log}\,{\Phi}_{-} = 0
\label{eq:windarg}
\eeq
Hence
\beq
\oint_{\gamma} \, {\rm log}\,{\Phi}_{+}\, \frac{dz}{z} = {\rm log}\,{\Phi}_{+}(0) = 
\oint_{\gamma} \, d{\rm log}\,{\Phi}_{-}\, \frac{dz}{z} = {\rm log}\,{\Phi}_{-}({\infty}) = 0
\label{eq:zravarg}
\eeq
and
\beq
y_{\si} (x, {\gamma}) \, = \, {\exp} \, \oint_{\gamma} \, {\rm log} \, \left[ (-z)^{-{\si}} {\sf P}(z, x) \right]\, \frac{dz}{z}
\label{eq:yifromaverage}
\eeq
is well-defined.
Let 
\beq
Z_{x} = {\rm the\ set\ of\ zeroes}\, {\rm of}\, {\sf P}({\cdot}, x)\, , \ {\CalD}_{x} = {\BC}^{\times} \backslash Z_{x}
\eeq
The points of ${\CalC}^{\rm spec}_{\si}$ are the equivalence classes
of the pairs $(x,{\gamma})$, where $(x,{\gamma}) \sim (x', {\gamma}')$ iff
$x = x'$ and ${\gamma}$ is homologous to $\gamma'$ on ${\CalD}_{x}$. More formally, 
\begin{multline}
{\CalC}^{\rm spec}_{\si} = 
\left\{ \, (x, [{\gamma}]) \, | \,  [{\gamma}] \in H_{1}\left(  {\CalD}_{x} \, , {\BZ} \right) \cap 
{\omega}^{\perp}
 \, \right\} \, , \\
 {\omega} \, =  \, d{\rm log}\,  \left[ (-z)^{-{\si}} {\sf P}(z, x) \right] \in H^{1}\left(  {\CalD}_{x} \, , 2{\pi}{\ii}{\BZ} \right) \ .
\label{eq:ispeccrve}
\end{multline}
The fact that ${\CalC}^{\rm spec}_{\si}$ is algebraic is not obvious in this description. Its advantage is the applicability to the ${\hat A}_{r}$-case. 

The ${\hat A}_{r}$-case can be now understood in a similar fashion, except
that the projection ${\pi}_{x}$ from the curve defined by the Eq. \eqref{eq:speccahr}
is ${\infty} : 1$. We model the fibers of the projection ${\pi}_{x}$ on 
${\BZ}$. The $i$'th spectral curve is modeled on the quotient $S({\BZ})/S(\{ \, n\, | \, n \leq i \}) \times S( \{ n\, |\, n > i \})$. To make this a tangible definition we use the formulae \eqref{eq:ispeccrve}, \eqref{eq:yifromaverage} with ${\sf P}(z,x)$ given by \eqref{eq:affprod}, \eqref{eq:ncj3}. Now there are infinitely many choices for the contour $\gamma$, so the $i$'th fundamental spectral curve is analytic, but most likely not algebraic.

\end{enumerate}

\subsection{Turning on the higher times}

Now let us turn the times back on. The Eqs. \eqref{eq:ncj},\eqref{eq:speccurvf}
hold, but we cannot use them efficiently, as we only know the characters $\chi_{\si}(x)$ are entire functions, but we cannot effectively control them because of the essential singularity at $x = \infty$ due to the $e^{\psi_{i}(x)}$ factors. 

The seemingly hopeless situation is saved by the following observation. The times $\bt$ are formal parameters. The curves \eqref{eq:speccahr} or \eqref{eq:speccar}  can be compactified to rational curves sitting in ${\BC\BP}^{1}_{x} \times {\CalE}_{\qe}$ or ${\BC\BP}^{1}_{x} \times {\BC\BP}^{1}_{z}$, respectively. The deformation of the problem by the higher times deforms these curves, preserving their general structure. In the ${\hat A}_{r}$-case we expect the curve ${\CalC}^{\rm spec}$ given by \eqref{eq:speccahr} to deform to $\widetilde{{\CalC}^{\rm spec}}^{\circ}$
\beq
x = {\sf A} + \sum_{{\si}=1}^{\sf r+1} (A_{\sf i-1} - A_{\si}) {\zeta}_{Q}(z/Z_{\si}) 
\label{eq:speccahrdef}
\eeq
with ${\sf A} = A_{\sf 0} - \frac{\ve_3}{2}$, ${\bA} = \left( A_{\si} \right), {\bZ} = \left( Z_{\si} \right)$, ${\si} \in {\BZ}$ obeying
\beq
A_{\si + {\sf r+1}} = A_{\si} - {\ve}_{3}\, , \ Z_{\si + {\sf r+1}} = Q Z_{\si}
\label{eq:defaq}
\eeq
with $\zeta_{Q}$ given by the Eq. \eqref{eq:zeta} with $\qe$ replaced by $Q$. We can impose the center-of-mass normalization on $Z_{\si}$'s:
\beq
\prod_{\sf i = 1}^{\sf r+1} Z_{\si} = 1\, , 
\eeq
mimicking \eqref{eq:cm}. By adding the points $p_{i} = (x,z)_{i} = ({\infty}, Z_{i}) \in {\BC\BP}^{1}_{x} \times {\BC}^{\times}$ for all $i \in {\BZ}$ to $\widetilde{{\CalC}^{\rm spec}}^{\circ}$ we get the curve $\widetilde{{\CalC}^{\rm spec}}$ with 
$\BZ$-action.

In the $A_r$-case the curve \eqref{eq:speccar} deforms to $\widetilde{{\CalC}^{\rm spec}}^{\circ}$ given by
\begin{multline}
x = m_{-} + \sum_{{\si}=1}^{\sf r+1}   \frac{Z_{\si}(A_{\sf i-1} - A_{\si})}{z-Z_{\si}}\, , \\
x \frac{dz}{z} = dz \left(  \frac{m_{+}}{z} + \sum_{{\si}=1}^{\sf r+1} \frac{A_{\sf i-1}-A_{\si}}{z-Z_{\si}}  \right) 
\label{eq:speccardef}
\end{multline}
with $A_{\sf r+1} = m_{+}, A_{\sf 0} = m_{-}$ undeformed.  By adding the points $p_{i} = (x,z)_{\si} = ({\infty}, Z_{\si}) \in {\BC\BP}^{1}_{x} \times {\BC\BP}^{1}_{z}$ for  ${\si} = 1, \ldots, r+1$,  as well as $p_{\sf 0} = (m_{-}, {\infty})$ and $p_{\sf r+2} =  (m_{+}, 0)$ to $\widetilde{{\CalC}^{\rm spec}}^{\circ}$ we get the curve $\widetilde{{\CalC}^{\rm spec}}$. 

The deformed data ${\bA}, {\bZ}$ is 
 to be found from the requirement:
 there is an entire function $\eig$ on $\widetilde{{\CalC}^{\rm spec}}$ with the asymptotics
 given by \eqref{eq:eigxas} near $z \to Z_{i}$, $i \in {\BZ}$, in the ${\hat A}_{r}$-case, and by 
 \eqref{eq:eigxasf} near $z \to Z_{\si}$, ${\si} = 1, \ldots, r+1$, in the $A_{r}$-case. 
 
 To find such a function and to find the constraints on $\bA, \bZ$ we introduce a sequence of 
 meromorphic functions ${\Omega}_{{\si}, k}$, $k \geq 1$, ${\si} = 1, \ldots, r+1$, on $\widetilde{{\CalC}^{\rm spec}}$, such that:
 \begin{enumerate}
 
 \item
 
 ${\Omega}_{{\si}, k}$ is holomorphic outside $z =  Z_{\si} Q^{\BZ}$, in the ${\hat A}_{r}$-case, and outside $z = Z_{\si}$ in the $A_r$-case. 
 
 \item
 
 Near $z = Z_{\si} Q^{p}$, ${\si} = 1, \ldots, r+1$, $p \in {\BZ}$, the function ${\Omega}_{{\si},k}$
 has the Laurent expansion:
 \beq
  {\Omega}_{{\si},k} = (x+p {\ve}_{3})^{k} + p {\varpi}_{{\si},k} + o(x^{-1})
  \eeq
  where ${\varpi}_{{\si}, k}$ are some constants, in the ${\hat A}_{r}$-case; and has the Laurent expansion:
  \begin{equation}
      \Omega_{i,k} = x^k + o(x^{-1})
  \end{equation}
  near $z = Z_{\si},\, \, {\si} = 1, \ldots, r+1$, in the $A_r$-case. 
 \end{enumerate}
For example, in the $A_r$-case:
\beq
\begin{aligned}
& {\Omega}_{{\si}, 1}(z) = 
\frac{A_{\si}- A_{\sf i-1}}{z/Z_{\si}-1} + m_{+} + \sum_{{\sj}\neq{\si}}^{\sf r+1} \frac{A_{\sj} - A_{\sf j-1}}{Z_{\si}/Z_{\sj}-1} \, , \\
& {\Omega}_{{\si},2}(z) = {\Omega}_{{\si}, 1}(z)^2 - 2 \left( A_{\si}- A_{\sf i-1} \right) Z_{i} \sum_{{\sj} \neq {\si}} \frac{\left( A_{\sj}- A_{\sf j-1} \right) Z_{j} }{\left( Z_{\si} - Z_{\sj}\right)^{2}} \, . \\
\end{aligned}
\eeq
Similarly, using
\beq
{\zeta}_{Q}(z) = \frac{1}{z-1} + \frac{1}{2} + {\sigma}_{1}(Q) (z-1) + \ldots\, , \ z\to 1 
\eeq
where
\beq
{\sigma}_{1}(Q) : = - 2\sum_{n=1}^{\infty} \frac{Q^{n}}{(1-Q^n)^2} 
\eeq
we compute in the ${\hat A}_{r}$-case: 
\begin{multline}
{\Omega}_{{\si}, 1}(z) = \left( A_{\sf i-1} - A_{\si} \right) {\zeta}_{Q}(z/Z_{\si})   + {\omega}_{\si} \, , \\
{\omega}_{\si} =  {\sf A} +  \sum_{{\sj} \neq {\si}} \left( A_{\sf j-1} - A_{\sj} \right) {\zeta}_Q (Z_{\si}/Z_{\sj}) \ , \\
 {\varpi}_{{\si},1} = A_{\si} - A_{{\si}-1} \ , 
\label{eq:h1aff}
\end{multline}
and
\begin{multline}
{\Omega}_{{\si},2}(z) \, =\,  (A_{\sf i-1} - A_{\si})^{2} \left( {\wp}_{Q}(z/Z_{\si}) + 3 {\sigma}_{1}(Q) \right) \\
+ 2 (A_{\sf i-1} - A_{\si}) {\omega}_{\si} \, {\zeta}_{Q}(z/Z_{\si})  \\
- 2 \left( A_{\si}- A_{\sf i-1} \right) \sum_{{\sj} \neq {\si}} \left( A_{\sj}- A_{\sf j-1} \right) {\wp}_{Q} \left( Z_{\si} /Z_{\sj} \right) \, , \\
{\varpi}_{{\si},2} = 2 (A_{\si} - A_{\sf i-1}) {\omega}_{\si}
\label{eq:h2aff}
\end{multline}
where
\begin{multline}
{\wp}_{Q}(z) = - z \frac{d}{dz} {\zeta}_{Q}(z) = \sum_{p \in {\BZ}}^{\infty} \frac{z Q^{p}}{(z- Q^{p})^{2}}   \, , \\
{\wp}_{Q}(z)\, = \, {\wp}_{Q}(z^{-1})  \, = \, 
{\wp}_{Q}(Q z)  \, . 
\label{eq:wp}
\end{multline}
The existence and uniqueness of the series of functions ${\Omega}_{{\si},k}(z)$
and constants ${\varpi}_{{\si},k}$ is easy to establish. 

Now define the function on $\widetilde{{\CalC}^{\rm spec}}$
\beq
{\eig} \, = \, z \, {\exp}\, \sum_{k=1}^{\infty} \sum_{{\si}=1}^{r+1} {\xi}_{{\si},k} {\Omega}_{{\si},k} (z)  \, , 
\label{eq:Zfun}
\eeq
regular outside $Z_{\sf 1}, \ldots , Z_{\sf r+1}$ ($\times Q^{\BZ}$ in the ${\hat A}_{r}$-case). Viewed as a multi-valued function of $x$, 
it has: 
\begin{enumerate}

\item
in the $A_r$-case, $r+1$ branches near $x = \infty$:
\beq
{\eig}_{\si}(x) = Z_{\si}\left( 1 + \frac{A_{\si}-A_{\sf i-1}}{x} + O(x^{-2}) \right) \, {\exp} \, \left( \sum_{k=1}^{\infty} {\xi}_{i,k} x^{k} \right) + O(x^{-1}) 
\eeq
which we identify with 
\beq
z_{\si}\, \frac{y_{\si}(x)}{y_{\sf i-1}(x)}\, \, {\exp} \, \left( \sum_{k=1}^{\infty} {\xi}_{{\si},k} x^{k} \right)\, ;  
\label{eq:funz1}
\eeq

\item
in the ${\hat A}_r$-case, an infinite set of branches near $x = \infty$, labelled by $({\si},p)$, 
${\si} = 1, \ldots, r+1$, $p \in {\BZ}$, or simply by $i = {\si} + p(r+1) \in {\BZ}$, 
\beq
{\eig}_{\si, p}(x) = Z_{\si} Q^{p}\left( 1 + \frac{A_{\si}-A_{\sf i-1}}{x} + O(x^{-2}) \right) \, {\exp} \, \left( \sum_{k=1}^{\infty} {\xi}_{i,k} (x+{\ve}_{3}p)^{k} \right) + O(x^{-1}) 
\eeq
which we identify with 
\beq
z_{\si}\, {\qe}^{p}\, \frac{y_{\si}(x+{\ve}_{3}p)}{y_{\sf i-1}(x+{\ve}_{3}p)}\, \, {\exp} \, \left( \sum_{k=1}^{\infty} {\xi}_{{\si},k} (x+{\ve}_{3}p)^{k} \right)\, ;  
\label{eq:funz2}
\eeq
\end{enumerate}
It remains to fix the \emph{string} \cite{KricheverW} equations needed  to fix the $2r+3$( $2r+4$) parameters  
$A_{\si}, Z_{\si}$ (and $Q$)
in the $A_r$ (${\hat A}_r$)-cases. These are obtained by matching the $x^{0}, x^{-1}$ terms in the large $x$ expansions of \eqref{eq:Zfun} and \eqref{eq:funz1}, \eqref{eq:funz2}: 
\beq
z_{\si} = Z_{\si} \, {\exp} \, \sum_{k=1}^{\infty} \sum_{{\sj}\neq {\si}} {\xi}_{{\sj},k} {\Omega}_{{\sj},k} (Z_{\si}) \, , \quad {\si} = {\sf 1}, \ldots, {\sf r+1}
\label{eq:zzeq}
\eeq
and
\begin{multline}
a_{\si}- a_{{\si}-1} = \frac{1}{2\pi\ii} \oint_{C_{\si}} x \frac{d{\eig}}{\eig} \\
=  A_{\si}-A_{\sf i-1} +  \sum_{k=1}^{\infty} \left( {\xi}_{{\si},k} {\tilde\Omega}_{{\si},k}  -  \left( A_{\si}-A_{\sf i-1}  \right) \sum_{{\sj} \neq {\si}} {\xi}_{{\sj},k} \, Z_{\si} {\Omega}_{{\sj}, k}^{\prime} (Z_{\si}) \right) \ , 
 \label{eq:stringeq}
 \end{multline}
 where
\beq
{\tilde\Omega}_{{\si},k} = \frac{1}{2\pi\ii} \oint_{C_{\si}} x d {\Omega}_{{\si},k}
\eeq
In the $A_r$ case we supplement \eqref{eq:stringeq} by
 \beq
 A_{\sf 0} = a_{\sf 0}\, ,  \qquad A_{\sf r+2} = a_{\sf r+2}\, ,  
\eeq

\subsection{Whitham hierarchy}

The Eqs. \eqref{eq:zzeq}, \eqref{eq:stringeq} produce implicit solution
to the infinite-dimensional dispersionless hierarchy of commuting
flows on the space of curves \eqref{eq:speccardef}, \eqref{eq:speccahrdef}, which carries
the symplectic structure
\beq
\boldsymbol{\omega} = \sum_{{\si}=1}^{\sf r+1} dp_{\si} \wedge \frac{dZ_{\si}}{Z_{\si}}
\eeq
which should be viewed as the result of symplectic reduction
of $( T^{*}{\BC}^{\times} )^{r+1}$ by the diagonal ${\BC}^{\times}$-action. 
The commutativity is obvious from the construction, as in the original (microscopic) problem the times $t_{i,k}$ could be moved independently. The flows are actually Hamiltonian  
in agreement with the general construction \cite{KricheverW} of I.~Krichever. The special nature of our hierarchy (it is a generalization of dispersionless Toda lattice found in \cite{LMN,MN})
is its microscopic origin. We shall present the details of the Hamiltonian structure of our hierarchy elsewhere.

\section{Higher spaces: elliptic cohomology}

There is yet another generalization of Vershik-Kerov problem, 
motivated \cite{BLN} by the studies of six dimensional gauge theory compactified on a two dimensional torus $T^2$. The theory depends on the choice of Euclidean metric on $T^2$ and a choice of $B$-field on $T^2$, which are conventionally packaged in two complex moduli, $\sigma$ and $\rho$. The $\sigma$-modulus parametrizes the complex structure of $T^2$ while $\rho$ captures the area, measured in units of inverse squared gauge coupling and the period of the $B$-field. Essentially the $\rho$ parameter becomes the complexified gauge coupling of the four dimensional gauge theory, so that the instanton counting parameter is 
\beq
{\qe} = e^{2\pi \ii \rho}
\eeq
The $\sigma$ parameter is encoded in  
\beq
{\pe} = e^{2\pi\ii \sigma}
\eeq
As a complex manifold, $T^2$ is identified with the elliptic curve ${\CalE}_{\pe} = {\BC}^{\times}/{\pe}^{\BZ}$. As gauge theories in dimension higher than four require ultraviolet completion, let us start with the IIB superstring background. 
Ten dimensional string background consists of the total space $\sf Y$ of a direct sum of four holomorphic degree zero line bundles $L_{a}$ over $T^2$, viewed as the elliptic curve ${\CalE}_{\pe} = {\BC}^{\times}/{\pe}^{\BZ}$. 
The bundles $L_a$ are parametrized by the points $q_a \in Jac({\CalE}_{\pe}) \approx {\CalE}_{\pe}$, 
and are required to obey
\beq
L_{1} \otimes L_{2} \otimes L_{3} \otimes L_{4} = {\CalO}_{{\CalE}_{\pe}}\, , \ q_{1}q_{2}q_{3}q_{4} = 1
\label{eq:trivdet}
\eeq
We denote by $\underline{q} = (q_{1},q_{2},q_{3},q_{4})$ the ordered $4$-tuple of points on ${\CalE}_{\pe}$ which sum up to $1$ in the sense of the group law on $\CalE_{\pe}$. The $IIB$ closed
string background is characterized by $\underline{q}$ up to an $S(4)$-permutation. The transformations of $\underline{q}$ 
\beq
q_{a} \mapsto q_{a} {\pe}^{n_{a}} \, , \ n_{a} \in {\BZ}\, , \qquad a = 1, 2, 3 , 4
\label{eq:gauge6d}
\eeq
with $n_1 + n_2 + n_3 + n_4 = 0$, are the isometries of $\sf Y$. Thus, $\sf Y$ is characterized by the $SU(4)$ flat connection $[{\underline{q}}]$, i.e. $\underline{q}$ modulo \eqref{eq:gauge6d} and the $S(4)$-permutations. 

In addition, our theory includes a $D5$-brane, wrapping the total space of the rank two vector bundle
$L_{1} \oplus L_{2}$ over ${\CalE}_{\pe}$. Its low energy configurations are described by a $U(1)$
vector multiplet of ${\CalN}=(1,1)$ supersymmetry, containing four real scalars, describing sections of  the line bundles $L_3$ and $L_4$. The $q_3, q_4$ twist parameters effectively make these scalars (and their ${\CalN}=1$ superpartners) massive. 

In summary, our theory is characterized by a choice of two elliptic curves ${\CalE}_{\qe}, {\CalE}_{\pe}$, and what looks like a semi-stable holomorphic $S(GL(2)\times GL(2))$-bundle on ${\CalE}_{\pe}$, i.e. a quadruple $\underline{q}$ obeying \eqref{eq:trivdet} modulo the equivalence relation \eqref{eq:gauge6d} and the $S(2)\times S(2)$ permutations. As we shall see below the intricate structure of anomalies
of the six dimensional theory leads to a more interesting geometry of the space of parameters.  

The measures \eqref{eq:macroplanch},\eqref{eq:plancht} depend on the geometry of $\lambda$ in rational way. The elliptic generalizations  result from calculations of various twisted elliptic genera, $\chi_y$-genera etc. of the Hilbert scheme of $N$ points on ${\BC}^{2}$, a.k.a. the moduli space of $U(1)$ instantons on noncommutative ${\BR}^{4}$, which is a particular case of the twisted Witten index of a ${\CalN}=(1,1)$ supersymmetric theory in six dimensions, compactified on a circle. According to the \cite{BLN} program (see also \cite{Hollowood:2003cv,Braden:2003gv}), the measure \eqref{eq:macroah0} is generalized to
the equivariant localization of elliptic genus, given by
\beq
{\mathscr M}_{\qe; \underline{q}; {\pe}} [{\lambda}] = \frac{1}{{\CalZ}({\qe}; \underline{q}; {\pe})} \, {\qe}^{|{\lambda}|} \, 
\prod_{{\square}\in {\lambda}} {\Upsilon}\left( q_{1}^{-{\sf l}_{\square}} q_{2}^{{\sf a}_{\square}+1}  ; q_{12}, q_{3}; {\pe}\right)  
\label{eq:ellmacroah0}
\eeq
where the $\Upsilon$-function is defined though the odd theta 
\begin{multline}
{\vt}(z; {\pe}) \,=\,  - {\vt}( z^{-1}; {\pe}) \, \\
= \, \sum_{r \in \BZ+\frac 12} (-1)^{r-\frac 12} \, {\pe}^{\frac{r^2}{2}} z^{r} \,
= \, z^{\frac 12} {\pe}^{\frac 18} {\phi}({\pe}) {\varphi}_{+}(z; {\pe}) {\varphi}_{-}(z; {\pe}) \, , \\
{\varphi}_{+}(z; {\qe}) = \prod_{n=1}^{\infty} (1-{\qe}^{n} z ) \, , \ 
{\varphi}_{-}(z; {\qe})  = \prod_{n=0}^{\infty} ( 1 - {\qe}^{n} z^{-1} ) 
\ , 
\label{eq:oddtheta}
\end{multline}
via
\beq
{\Upsilon}(z; t_{1}, t_{2}; {\pe}) = \frac{{\vt}(t_{2} z; {\pe}){\vt}(t_{1}t_{2} z^{-1}; {\pe})}{{\vt}(z; {\pe}) {\vt}(t_{1} z^{-1}; {\pe})}
\label{eq:upsfun}\eeq
If a variable $u$ transforms under \eqref{eq:gauge6d} as $u \mapsto u {\pe}^{n_{u}}$, then\footnote{using
\beq
{\vt}( {\pe}^{n} e^{2\pi\ii m} z ; {\pe} ) \, = \, (-1)^{m+n} \, {\pe}^{-\frac{n^2}{2}} z^{-n} \, {\vt}(z; {\pe}) \ . 
\label{eq:thtr}
\eeq}
\begin{multline}
{\Upsilon}(u; q_{12}, q_{3}; {\pe}) \mapsto
{\Upsilon}(u; q_{12}, q_{3}; {\pe})  \times {\Sigma}({\underline{q}}, \underline{n} )\, , \\
{\Sigma}({\underline{q}}, \underline{n} ) = q_{3}^{n_{4}} q_{4}^{n_{3}}  {\pe}^{n_{3}n_{4}}  = 
q_{12}^{-n_{3}} q_{3}^{-2n_{3}-n_{1}-n_{2}} {\pe}^{-n_{3}(n_{1}+n_{2}+n_{3})} \ .
\end{multline}
The measure \eqref{eq:ellmacroah0} is invariant if
the instanton coupling  $\qe$ transforms non-trivially
\beq
{\qe} \mapsto \, {\Sigma}({\underline{q}}, \underline{n} )^{-1}\, {\qe}
\label{eq:gs1}
\eeq
Note the manifest $S(2) \times S(2)$ invariance of the $\Sigma$-cocycle. 

Leaving the physics \cite{NP} interpretation of \eqref{eq:gs1} aside\footnote{In modern language, it has to do with the mixed anomaly between the topological $1$-form symmetry and the flavor $0$-form symmetry in six dimensions}, let us solve the limit shape problem, i.e. find the $q_1, q_2 \to 1$ limit while keeping $\qe, q_3, \pe$ finite. 

The $Y$-observable in the 6d theory is defined by the obvious elliptic analogue of \eqref{eq:yobs}:
\begin{equation}
    {\mathscr Y}(x) [{\lambda}]  = {\vt}(x; {\pe}) \prod_{\square \in \lambda} \frac{{\vt} \left( x {\sf Q}_{\square}^{-1} (q_1,q_2) q_{1}^{-1}  ; {\pe} \right) {\vt} \left( x {\sf Q}_{\square}^{-1} (q_1,q_2) q_{2}^{-1}  ; {\pe} \right)}{{\vt} \left( x {\sf Q}^{-1}_{\square} (q_1,q_2) ; {\pe} \right) {\vt} \left( x {\sf Q}^{-1}_{\square} (q_1,q_2) q_{1}^{-1} q_{2}^{-1} ; {\pe} \right)}
    \label{eq:Yell}
\end{equation}
For any $\lambda$, ${\mathscr Y}(x)[{\lambda}]$ is a meromorphic section of a degree $1$ line bundle $L$ over ${\CalE}_{\pe}$. The isomorphism class of $L$ is $\lambda$-independent, making possible the setting of the compactness theorem of \cite{BPSCFT1, BPSCFT2} which implies that the expectation values of the elliptic analogue of \eqref{eq:xarh}
\begin{multline}
{\mathscr X}^{{\hat A}_{0}} (x) =  
\, {\mathscr Y}(x q_{12}) 
 \times \\
\sum_{\nu \in {\Lambda}} {\mathscr M}_{\bq; q_{3}, q_{4}, q_{1}, q_{2}; {\pe}} [ {\nu}] \prod_{{\square} = (i,j) \in {\nu}} \frac{{\mathscr Y} \left( x {\sf Q}_{\square} (q_{3}, q_{4}) q_{3}^{-1} \right) }{{\mathscr Y}\left( x {\sf Q}_{\square} (q_{3}, q_{4})  \right)} \frac{{\mathscr Y} \left(x {\sf Q}_{\square} (q_{3}, q_{4}) q_{4}^{-1} \right)}{{\mathscr Y} \left(x {\sf Q}_{\square} (q_{3}, q_{4})  q_{12} \right)}
\label{eq:ellxarh}
\end{multline}
is a holomorphic section of $L$, i.e. proportional to ${\vt}(x; {\pe})$. Taking the $q_{1}, q_{2} \to 1$ limit, we conclude (cf. \eqref{eq:ellmacroah0}): 
\beq
{\chi}^{6d}(x)  : = \left\langle {\mathscr X}^{{\hat A}_{0}} (x) \right\rangle_{{\qe};  q_{1}, q_{2}, q_{3}, q_{4}; {\pe}} \, \biggr\vert_{q_{1},q_{2}\to 1}  = c \cdot {\vt}(x; {\pe}) \, , \label{eq:6dchar} \ ,  
\eeq
where the prefactor $c = c({\qe}, {\pe}, q_{3})$ can be computed by evaluating both sides at $x = 1$. 
Again, we perform a theta-transform to get the infinite product formula:
\begin{multline}
\label{thetatransform6d}
    {\Phi}(z, x ; {\qe}, {\pe}, m) = c({\qe}, {\pe}, q_3) \sum_{M \in {\BZ}} (-z)^M {\qe}^{\frac{M(M-1)}{2}} {\vt}(x  q_{3}^{M} ; {\pe})  \\
    = \prod_{n=1}^{\infty} \left( 1 - z^{-1} {\qe}^{n} \frac{{\mathscr Y}(x q_{3}^{n})}{{\mathscr Y}(x q_{3}^{n-1})} \right) \, {\mathscr Y}(x) \, \prod_{n=0}^{\infty} \left( 1 - z {\qe}^{n} \frac{{\mathscr Y}(x q_{3}^{-n-1})}{{\mathscr Y}(x q_{3}^{-n})} \right) 
\end{multline}
The vanishing locus of $\Phi$ is a curve ${\CalC}_{6d}$ in the $(z,x)$-space, with the branches
\beq
{\eig}_{n}(x) = {\qe}^{n} \frac{{\mathscr Y}(x q_{3}^{n})}{{\mathscr Y}(x q_{3}^{n-1})}\, , \ n \in {\BZ}\, , 
\label{eq:pbranch}
\eeq
containing all the information about the limit shape, as in the rational, four dimensional, case. Now, ${\Phi}$ is trivially seen to be a genus two theta function
\beq
{\Phi} \propto {\Theta} \left( x, z \, \vert \, {\CalT} \right) 
\eeq
with a period matrix
\beq
{\CalT} = \left( \begin{matrix} {\sigma} & m \\ m & {\tau} \end{matrix} \right)
\label{eq:period}
\eeq
where $m = \frac{1}{2\pi\ii} {\rm log}(q_{3})$. 
Thus ${\CalC}_{6d}$ is a genus two curve, the equation $\Phi = 0$ being the theta divisor in the abelian variety Jac$({\CalC}_{6d})$, describing the embedding of the curve into its Jacobian by the Abel-Jacobi map. The transformation
\eqref{eq:gs1} is just one class of the $Sp(4, {\BZ})$ modular transformations.

\section{Conclusions and further directions}

In this paper we have constructed the limit shapes of two infinite sequences of multi-parametric generalizations of the Plancherel measure studied in \cite{VK, LS}. These generalizations come from the enumerative geometry of moduli spaces of rank one sheaves on complex surfaces embedded into a Calabi-Yau fourfold, perhaps with a transverse Kleinian singularity of $A$-type. We made contact
with Whitham hierarchies \cite{KricheverW}, generalizing and clarifying some of the earlier work on the subject \cite{Gorsky:1998rp, LMN, MN}. 
We clarified the notion of cameral and spectral curves for the limit shape problem, 
a novel notion in the affine case. 
The tools we employed: the generalized Jacobi identity found in \cite{GNncJ}, the $\theta$-transforms etc. will be used in \cite{GNtoapp} to construct the Lax operators, their eigenvectors, as well as isomonodromic connections and their horizontal sections, associated with rank $N$ vector bundles on genus zero and one curves with punctures. 

In our paper we illustrated the power of our methods at the example of a double-elliptic generalization of Vershik-Kerov problem, showing a non-trivial emergent geometry of genus two abelian varieties. 

In the forthcoming work \cite{GNtoapp} (partly previewed in \cite{G}) we shall study the higher rank sheaves (non-abelian gauge theories), 
parabolic sheaves (surface defects), and affine quiver generalizations of the elliptic cohomology problem (six dimensional cyclic quiver theories). The higher times deformations we explored
in rank one case will eventually become related to the flows of algebraic integrable systems long conjectured \cite{Gorsky:1995zq, DW} to be connected to the low-energy physics of supersymmetric gauge theories in four (macroscopic) dimensions. 

The Unity of Mathematics (and its deep connections to theoretical physics) was always a point of fascination and a source of inspiration for Anatoly Moiseevich.

\end{document}